\documentclass[preprint]{ptephy_v1}

\preprintnumber{YITP-16-128, MPP-2016-335, OU-HET-917} 

\usepackage{graphics} 

\newcommand{\rme}{{\rm e}}
\newcommand{\rmd}{{\rm d}}
\newcommand{\rmD}{{\rm D}}

\begin{document}

\title{Flow equation for  the scalar model in the large $N$ expansion and its applications}


\author{Sinya Aoki}
\affil{Center for Gravitational Physics, Yukawa Institute for Theoretical Physics, Kyoto University, 
Kitashirakawa Oiwakechou, Sakyo-ku, Kyoto 606-8502, Japan
 \email{saoki@yukawa.kyoto-u.ac.jp}}

\author{Janos Balog}
\affil{Institute for Particle and Nuclear Physics, 
Wigner Research Centre for Physics, 
MTA Lend\"ulet Holographic QFT Group,
1525 Budapest 114, P.O.B.\ 49, Hungary \email{balog.janos@wigner.mta.hu}}

\author{Tetsuya Onogi} 
\affil{Department of Physics, Osaka University, Toyonaka, Osaka 560-0043, Japan \email{onogi@phys.sci.osaka-u.ac.jp}}

\author{Peter Weisz}
\affil{Max-Planck-Institut f\"ur Physik, 80805 Munich, Germany \email{pew@mpp.mpg.de}}


\begin{abstract}
We study the flow equation of the O($N$) $\varphi^4$ model in $d$ dimensions at the next-to-leading order (NLO) in the $1/N$ expansion.
Using the Schwinger-Dyson equation, we derive 2-pt and 4-pt functions of flowed fields. 
As the first application of the NLO calculations, 
we study  the running coupling defined from the connected 4-pt function of flowed fields
in the $d+1$ dimensional theory. We show in particular that this running coupling
has not only the UV fixed point but also  an IR fixed point (Wilson-Fisher fixed point) in  the 3 dimensional massless scalar theory.   
As the second application, we calculate the NLO correction to the induced metric in $d+1$ dimensions with $d=3$
in the massless limit. While the induced metric describes a 4-dimensional Euclidean Anti-de-Sitter (AdS)  space at the leading order as shown in the previous paper,  the NLO corrections make the space asymptotically  AdS  only in UV and IR limits. Remarkably, while the AdS radius  does not receive  a NLO correction in the UV limit, the AdS radius decreases at the NLO  in the IR limit, which corresponds to the Wilson-Fisher fixed point in the original scalar model in 3 dimensions.
\end{abstract}

\subjectindex{B30, B32,B35,B37}

\maketitle

\section{Introduction}
\label{sec:intro}
In the previous paper\cite{Aoki:2016ohw}, 
the present authors studied the proposal\cite{Aoki:2015dla} that a $d+1$ dimensional induced metric 
can be constructed from a $d$ dimensional field theory using gradient flow\cite{Narayanan:2006rf,Luscher:2010iy, Luscher:2009eq,Luscher:2013vga}, applying the method to  the O($N$) $\varphi^4$ model.
We have shown that in the large $N$ limit the induced metric becomes classical and describes Euclidean 
Anti-de-Sitter (AdS) space in both ultra-violet (UV) and infra-red (IR) 
limits of the flow direction.
The method proposed in Ref.~\cite{Aoki:2015dla} may provide an alternative way to understand 
the AdS/CFT (or more generally Gravity/Gauge theory) correspondence\cite{Maldacena:1997re},
and the result in Ref.~\cite{Aoki:2016ohw} might be related to the correspondence between  O($N$) vector models in $d$-dimensions and (generalized) gravity theories in $d+1$ dimensions\cite{Klebanov:2002ja}.

To further investigate a possible connection between Ref.~\cite{Aoki:2016ohw} 
and Ref.~\cite{Klebanov:2002ja} at the quantum level,
one must calculate, for example,  the anomalous dimension of the O($N$) invariant operator $\phi^2(x)$, which requires the next-to-leading order (NLO) of the $1/N$ expansion for the flow equation to evaluate necessary quantum corrections. Since  the method employed in Refs.~\cite{Aoki:2015dla,Aoki:2016ohw} is a specific one adopted for  the large $N$ limit, some systematic way to  solve the flow equation in the $1/N$ expansion is needed.  

In this paper, we employ the Schwinger-Dyson equation (SDE) to solve the flow equation in the $1/N$ expansion
for the O($N$) invariant $\varphi^4$ model in $d$ dimensions. Using this method we explicitly calculate the 2-pt and 4-pt functions at the NLO.

As the first application of the NLO calculations, we define a running coupling from
the connected 4-pt function of flowed fields, which  runs  with the flow time $t$ such that $t=0$ corresponds to the UV limit while $t=\infty$ is the IR limit. 
This property establishes that the flow equation can be interpreted as a renormalization group transformation.
In particular at $d=3$, we show that the running coupling so defined  has not only the asymptotic free UV fixed point but also  a Wilson-Fisher IR fixed point for the massless case. 

As the second application, we investigate the NLO correction to the induced metric in $3+1$ dimensions  from the massless scalar model in 3 dimensions.
In the massless limit, the whole 4-dimensional space becomes AdS at the leading order, as shown in Ref.~\cite{Aoki:2016ohw}. The NLO corrections give  a small perturbation to the metric, which makes the space asymptotically AdS in UV ($t=0$) and IR ($t=\infty$) limits only. A  remarkable thing is that, while the NLO corrections do not change
the AdS radius in the UV limit, the AdS radius is  reduced by the NLO correction in the IR limit, which corresponds to the Wilson-Fisher IR fixed point of the original theory. In other words, a nontrivial fixed point in the field theory  leads to   a change of the AdS radius  in the geometry at the NLO.
The induced metric at NLO describes a 4-dimensional space connecting one asymptotically AdS space at UV to an other  asymptotically AdS space at IR, which have different radii. 

This paper is organized as follows. 
In Sec.~\ref{sec:flow}, we first introduce the  O($N$) invariant $\varphi^4$ model in $d$ dimensions. We then formulate  the Schwinger-Dyson equation (SDE) for the flowed fields, and solve it 
to derive 2-pt and 4-pt functions of flowed fields at the NLO.
In Sec.~\ref{sec:coupling}, we define  a running  coupling  from the connected 4-pt function of flowed fields  and investigate its behavior as a function of the flow time $t$.
In Sec.~\ref{sec:metric}, we study the induced metric from the 3 dimensional massless scalar model at the NLO.
We finally give a summary of this paper in Sec.~\ref{sec:summary}.
We collect all technical details in appendices. 
In appendix~\ref{app:d-dim}, using the SDE,
we present results at the NLO in the $1/N$ expansion of the $d$ dimensional theory necessary for the main text.
We also perform the renormalization  of the $d$ dimensional theory at the NLO, and 
explicitly calculate renormalization constants for various $d$.
In appendix~\ref{app:SDE_flow}, we give detailed  derivations of solutions to the SDE for the flow fields at the NLO.
We explicitly evaluate 2-pt and 4-pt functions of the flowed  field in appendix~\ref{app:massless} while we
derive the induced metric in appendix~\ref{app:metric}, for the massless scalar theory in 3 dimensions.

\section{$1/N$ expansion of the flow equation in $d+1$ dimensions}
\label{sec:flow}
\subsection{Model in $d$ dimensions}
In this paper, we consider  the  $N$ component scalar $\varphi^4$ model in $d$ dimensions, 
defined by the action
\begin{eqnarray}
S(\mu^2,u) &=&  N \int \rmd^d x\,\left[ \frac{1}{2}\partial^k \varphi (x)\cdot \partial_k \varphi (x) +  \frac{\mu^2}{2}\varphi^2(x) +\frac{u}{4!} \left(\varphi^2(x)\right)^2
\right] ,
\label{eq:action}
\end{eqnarray}
where $\varphi^a(x)$ is an $N$ component scalar field, $(\ \cdot \ )$ 
indicates an inner product of $N$ component vectors such that 
$\varphi^2(x) \equiv \varphi(x)\cdot\varphi(x) =\sum_{a=1}^N \varphi^a(x)\varphi^a(x)$, 
$\mu^2$  is the bare scalar mass parameter, and $u$ is the coupling 
constant of the $\varphi^4$ interaction, whose canonical dimension is $4-d$.
While it is consistent to take $u$ as $N$ independent,
as will be seen later, the mass parameter $\mu^2$ is expanded as
\begin{eqnarray}
\mu^2 &=& \mu_0^2 + \frac{1}{N}\mu_1^2 + \cdots,
\end{eqnarray}
where $\mu_i^2$ is cut-off dependent in order to make the physical mass finite order by order in the  $1/N$ expansion.

This model describes the free massive scalar at $u=0$, while 
it is equivalent to the non-linear $\sigma$ model (NLSM) in the $u\rightarrow \infty$ limit, 
whose action is obtained from eq.~(\ref{eq:action}) as
\begin{eqnarray}
S(\lambda) &=& \frac{ N}{2\lambda} \int \rmd^d x\,\partial^k \sigma (x)\cdot \partial_k \sigma (x) , \qquad
\sigma^2(x) = 1, 
\label{eq:action_NLSM}
\end{eqnarray}
with the replacement
\begin{eqnarray}
\sigma^a(x)&=&\sqrt{\lambda} \varphi^a(x), \qquad \lambda 
=  \lim_{u\rightarrow\infty}  -\frac{u}{6\mu^2} . 
\end{eqnarray}

Some regularization which preserves O($N$) symmetry is assumed in  this paper,
so that we can always make formal manipulations without worrying about divergences.\footnote{We will call the infinite cutoff ($\Lambda\to\infty$) limit the 'continuum limit'.}
Calculations of 2-pt and 4-pt functions at the next-to-leading order (NLO) of the $1/N$ expansion in $d$ dimensions 
will be given in appendix~\ref{app:d-dim}.

\subsection{Flow equation in the $1/N$ expansion }
In this paper, we consider the flow equation, given by
\begin{eqnarray}
\frac{\partial}{\partial t} \phi^a(t,x) &=& -\left. \frac{1}{N} \frac{\delta S(\mu_f^2,u_f)}{\delta \varphi^a(x)}\right\vert_{\varphi\rightarrow\phi}
= \left(\Box -\mu_f^2\right)\phi^a(t,x) -\frac{u_f}{6}\phi^a(t,x)
\phi^2(t,x), 
\label{eq:flow} \\
\phi^a(0,x) &=&\varphi^a(x), \nonumber
\end{eqnarray}
where $\mu_f^2$ and $u_f$ can be different from $\mu^2$ and $u$ in 
the original $d$ dimensional theory.
As in the case of $d$ dimensions,  $u_f$ is kept fixed and $N$ independent, whereas $\mu^2_f$ is adjusted as 
\begin{equation}
\mu_f^2 = m_f^2 - \frac{u_f}{6} Z(m_f), \qquad Z(m_f) \equiv \int \rmD q \frac{1}{q^2+m_f^2}, \quad \rmD q \equiv \frac{\rmd^d q}{(2\pi)^d},  
\label{eq:Zm}
\end{equation}
where $m_f$ is  a renormalized mass.
The flow with $\mu_f=\mu$ and $u_f=u$ is called gradient flow, as it is 
given by the gradient of the original action. 

In the case of the free flow ($u_f=0$), the solution is easily given by
\begin{eqnarray}
\phi^a(t,x) &=& \rme^{t(\Box -\mu_f^2)} \varphi^a(x).
\end{eqnarray}
We therefore consider the interacting flow ($u_f\not=0$) hereafter unless otherwise stated.

The above flow equation leads to the Schwinger-Dyson equation (SDE)\cite{Aoki:2014dxa}  as
\begin{eqnarray}
\langle D^f_z \phi^a(z) {\cal O} \rangle &=& -\frac{u_f}{6} \langle \phi^a(z) \phi^2(z) {\cal O}\rangle,  
\quad D_z^f \equiv \frac{\partial}{\partial t} -(\Box -\mu_f^2), 
\end{eqnarray}
where $z=(t,x)$, ${\cal O}$ is an arbitrary operator and the expectation value $\langle{\cal O} \rangle$ should be calculated in $d$  dimensions as
\begin{eqnarray}
\langle {\cal O}(\varphi)\rangle &\equiv & \frac{1}{Z} \int \left[{\cal D}\varphi\right] {\cal O}(\varphi)\rme^{-S(\mu^2,u)}, \quad
Z = \int \left[{\cal D}\varphi\right] \rme^{-S(\mu^2,u)} .
\label{eq:vev_d}
\end{eqnarray}

If we take ${\cal O} =\displaystyle\prod_{i=1}^{2n-1} \phi^{a_i}(z_i)$ the SDE becomes 
\begin{eqnarray}
D_z^f \, \Gamma_{2n}^{aa_1\cdots a_{2n-1}}(z,z_1,\cdots,z_{2n-1}) &=&
-\frac{u_f}{6 N^2}\sum_b \Gamma_{2n+2}^{abba_1\cdots a_{2n-1}}(z,z,z,z_1,\cdots,z_{2n-1}),~~
\end{eqnarray}
where $\Gamma_n$ is the $n$-point function, defined by
\begin{eqnarray}
\Gamma_n^{a_1\cdots a_n}(z_1,\cdots,z_n) = N^{n-1}\langle \prod_{i=1}^n \phi^{a_i}(z_i)\rangle 
\equiv \Gamma_n[12\cdots n] ,
\end{eqnarray}
which is analogous to the $d$  dimensional  counterpart in eq.~(\ref{eq:G2n_d}).
We consider only the symmetric phase in this paper, where $\Gamma_{2n-1} = 0$ for all positive integers $n$. 

We consider the next-to-leading order of the $1/N$ expansion, so that the following two SDE's need to be considered.
\begin{eqnarray}
D_1^f\, \Gamma_2[12] &=& -\frac{u_f}{6N^2} \sum_b \Gamma_4[1bb2], \\
D_1^f\, \Gamma_4[1234] &=& 
-\frac{u_f}{6N^2} \sum_b \Gamma_6[1bb234],  
\end{eqnarray}
where $z_b=z_1$, so that  the sum over $b$ runs over the O($N$) index only.

The connected part of  4- and 6- pt functions are introduced as 
\begin{eqnarray}
\Gamma_4[1234] &=& K_4[1234] + N\left\{\Gamma_2[12] \Gamma_2[34] + \Gamma_2[13] \Gamma_2[24] + \Gamma_2[14] \Gamma_2[23]\right\}, \\
\Gamma_6[123456] &=& K_6[123456] + N\left\{\Gamma_2[12] K_4[3456] + \mbox{ 14 perms.} \right\} \nonumber \\
&+& N^2\left\{\Gamma_2[12] \Gamma_2[34]\Gamma_2]56] + \mbox{ 14 perms.} \right\} .
\end{eqnarray}
Furthermore  decompositions in O($N$) indices are given by
\begin{eqnarray}
\Gamma_2[12] &=& \delta^{a_1a_2}\Gamma(z_1,z_2), \\
K_4[1234] &=& \delta^{a_1a_2}\delta^{a_3a_4} K(z_1,z_2;z_3,z_4) +\mbox{ 2 perms.}, \\
K_6[123456] &=& \delta^{a_1a_2}\delta^{a_3a_4}\delta^{a_5a_6} H(z_1,z_2;z_3,z_4;z_5,z_6) +
\mbox{ 14 perms.} ,
\end{eqnarray}
where $\Gamma(z_1,z_2)$, $K(z_1,z_2;z_3,z_4)$ and $H(z_1,z_2;z_3,z_4;z_5,z_6)$ are invariant under the exchange of arguments such that $z_{2i-1}\leftrightarrow z_{2i}$ or $(z_{2i-1},z_{2i}) \leftrightarrow (z_{2j-1},z_{2j})$. 

By expanding $\Gamma$, $K$ and $H$ as
\begin{equation}
\Gamma =\sum_{i=0}^\infty \frac{\Gamma_i}{N^i}, \qquad
K=\sum_{i=0}^\infty \frac{K_i}{N^i}, \qquad
H=\sum_{i=0}^\infty \frac{H_i}{N^i}, 
\end{equation}
the above two SDE are reduced to
\begin{eqnarray}
D_1^f \Gamma_0(12) &=& -\frac{u_f}{6} \Gamma_0(12)\Gamma_0(11)
\label{eq:SDE_LO}
\end{eqnarray}
at the LO of the $1/N$ expansion, and
\begin{eqnarray}
D_1^f \Gamma_1(12) &=& -\frac{u_f}{6}\left[K_0(12;11) +\Gamma_0(12)\Gamma_1(11)+
\Gamma_1(12)\Gamma_0(11) +2\Gamma_0(12)\Gamma_0(11)\right], ~~~~~ 
\label{eq:SDE_G1}\\
D_1^f K_0(12;34) &=&  -\frac{u_f}{6}\left[\Gamma_0(12) K_0(11;34) +\Gamma_0(11) K_0(12;34) 
+2\Gamma_0(12)\Gamma_0(13) \Gamma_0(14)\right] ~~~~~~~~
\label{eq:SDE_K0}
\end{eqnarray}
at the NLO.

\subsection{Solutions to the flowed SDE at NLO}
The solutions to the SDE at NLO are summarized below. Details of calculations can be found in appendix~\ref{app:SDE_flow}.

At the NLO, the 2-pt function is given by
\begin{eqnarray}
\langle \phi^{a_1}(z_1) \phi^{a_2}(z_2) \rangle &=&\frac{\delta_{a_1a_2}}{N}
 \frac{Z(m_f)}{\sqrt{\zeta(t_1)\zeta(t_2)}}
  \int \rmD p \frac{\rme^{-p^2(t_1+t_2)}\rme^{i p(x_1-x_2)}}{p^2+m^2} 
\left[ 1+\frac{1}{N} G_1(t_1,t_2\vert p) \right], ~~~~~
\end{eqnarray}
where $\zeta(t)$ is defined in eq.~(\ref{eq:zeta}), and the NLO contribution $G_1(t_1,t_2\vert p)$
is given in appendix~\ref{app:SDE_flow_G1}.
In the continuum limit, $\zeta(t)$ approaches to $\zeta_0(t)$ and  is finite as long as $t> 0$,
where
\begin{eqnarray}
\zeta_0(t) &\equiv& \int \rmD p\frac{\rme^{-2p^2t}}{p^2+ m^2}=\frac{\rme^{2tm^2} m^{d-2}}{(4\pi)^{d/2}}\Gamma(1-d/2,2tm^2)
\label{eq:zeta0}
\end{eqnarray}
with the incomplete gamma function $\Gamma(a,x)$, while $Z(m_f)$ diverges at $d > 1$.

The leading contribution of the connected 4-pt function appearing at the NLO of the $1/N$ expansion can be obtained as
\begin{eqnarray}
\langle \phi^{a_1}(z_1) \phi^{a_2}(z_2) \phi^{a_3}(z_3)\phi^{a_4}(z_4)\rangle_c
&=& \frac{1}{N^3}\left[\delta_{a_1a_2}\delta_{a_3a_4}K_0(12;34) + \mbox{ 2 permutations} \right] ,
\label{eq:4pt_c}
\end{eqnarray}
where 
\begin{eqnarray}
K_0(12;34) &=& \int  \rmd P_4\ g(12;34\vert 12;34),
\quad \rmd P_4 \equiv \prod_{j=1}^4  \rmD p_j \sqrt{\frac{Z(m_f)}{\zeta(t_j)}} \frac{\rme^{ip_j x_j} \rme^{-p_j^2 t_j}}{p_j^2+m^2},
  \label{eq:K0g} 
\\
g(12;34\vert 12;34) &=& X(23\vert 12;34)+X(13\vert 21;34)+X(24\vert 12;43)+X(14\vert 21;43)\nonumber \\
&+&Y(2\vert 12;34) + Y(1\vert 21;34) +Y(3\vert 43;12) +Y(4\vert 34;12) \nonumber \\
&+& Z(\vert 12;34) .
\end{eqnarray}
 Here the variables to the left of the vertical
  line refer to flow times and those to the right refer to momenta.
Explicitly we have  in the continuum or NLSM limits
\begin{eqnarray}
X(t_1,t_2\vert 12;34) &=& \hat\delta (p_2^2+m^2)(p_3^2+m^2)\int_0^{t_1} \rmd s_1\, 
\int_0^{t_2} \rmd s_2\, \rme^{s_1(p_2^2-p_1^2)} \rme^{s_2(p_3-p_4^2)} \omega(s_1,s_2\vert p_{34}),~~~~~~ \\
Y(t\vert 21;34) &=& \hat\delta (p_1^2+m^2)\int_0^t \rmd s\, \rme^{s(p_1^2-p_2^2)}\psi(s\vert 34), \\
Z(\vert 12;34) &=& -\hat\delta \frac{2}{6/u+ B(0\vert p_{34})}, 
\end{eqnarray}
where $\hat\delta \equiv (2\pi)^d \delta(p_1+p_2+p_3+p_4)$, $p_{34} =p_3+p_4$, 
\begin{eqnarray}
B(t\vert Q) &=& \int \rmD q_1 \rmD q_2 \frac{\rme^{-t(q_1^2+q_2^2)}}{(q_1^2+m^2)(q_2^2+m^2)} (2\pi)^d\delta (q_{12}-Q), \quad q_{12}=q_1+q_2, 
\label{eq:Bt}
\end{eqnarray}
and thus $B(0\vert Q) = B(Q^2)$, defined in appendix~\ref{app:d-dim}.
Here
$\psi$ and $\omega$ satisfy
\begin{eqnarray}
\rho(t\vert 34) +\int_0^t \rmd s\, K(t,s\vert  p_{34}) \psi(s\vert 34)  &=& 0, \label{eq:psi} \\
\rho(t_1,t_2\vert Q)  - 2 \int_0^{t_1} \rmd s_1\, K(t_1,s_1\vert Q) \int_0^{t_2} \rmd s_2\, K(t_2,s_2\vert Q)\, \omega(s_1,s_2\vert Q)
&=& 0, 
\label{eq:omega}
\end{eqnarray}
where
\begin{eqnarray}
K(t,s\vert Q) &=& \int \rmD q_1 \rmD q_2\, (2\pi)^d \delta(q_{12}-Q) \frac{\rme^{-(t+s)q_1^2-(t-s)q_2^2}}{q_1^2+m^2} , 
\label{eq:Kts}\\
\rho(t\vert 34) &=&\rme^{- t(p_3^2+p_4^2)} -\frac{B(t\vert p_{34})}{6/u + B(0\vert p_{34})} , 
\label{eq:rho_t}\\
\rho(t_1,t_2\vert Q) &=& B(t_1+t_2\vert Q) -\frac{B(t_1\vert Q) B(t_2\vert Q)}{6/u +B(0\vert Q)} .
\label{eq:rho_ts}
\end{eqnarray}
The derivation of these results is given in appendix~\ref{app:SDE_flow}.

\section{Running coupling from flowed fields}
\label{sec:coupling}
\subsection{Definitions}
Using the connected 4-pt functions  $g \equiv \hat \delta \hat g$ for the flow fields given in eq.~(\ref{eq:4pt_c}),
we define the $t$-dependent dimensionless coupling as
\begin{eqnarray}
g(t) = -3 \hat g(t,t; t,t\vert \{ p\}_{\rm sym}) t^{2-d/2},  
\end{eqnarray}
where $\{ p\}_{\rm sym}$ is given by $p_i^2 t = 3\Delta/4$ ($i=1\sim 4$) and $p_{12}^2 t =p_{34}^2 t=\Delta$
($p_{ij}=p_i + p_j$), which is the symmetric point for $d > 2$, and $t^{2-d/2}$ is introduced to make the coupling dimensionless. Here $\Delta$ is an arbitrary dimensionless constant but we can take $\Delta = 1$ without loss of generality by the rescaling $t\rightarrow \Delta t$. 
Explicitly we have
\begin{eqnarray}
\hat  g(t,t; t,t\vert \{ p\}_{\rm sym}) &=& 4 \hat X(t,t\vert  \{ p\}_{\rm sym}) + 4 \hat Y(t\vert  \{ p\}_{\rm sym}) +  \hat Z(\vert \{ p\}_{\rm sym}),
\end{eqnarray}
where we remove $\hat\delta$ by defining ${\cal O} = \hat\delta \hat {\cal O}$ for ${\cal O} = g, X, Y, Z$, and 
\begin{eqnarray}
\hat X(t_1,t_2\vert 12;34) &=& (p_2^2+m^2)(p_3^2+m^2)\int_0^{t_1} \rmd s_1\int_0^{t_2} \rmd s_2\,
\rme^{s_1(p_2^2-p_1^2)}\rme^{s_2(p_3^2-p_4^2)}\omega(s_1,s_2\vert p_{34}),~~~~~~ \\
\hat Y(t\vert 12;34) &=& (p_2^2+m^2) \int_0^t \rmd s\, \rme^{s(p_2^2-p_1^2)}\psi(s\vert  34), \\
\hat Z(\vert 12;34) &=& -\frac{1}{3}\frac{u}{1+\frac{u}{6} B(0\vert p_{34})} .
\end{eqnarray}

\subsection{Free flow}
For simplicity, we first consider the free flow, where $\hat g(t,t;t,t\vert \{p\}_{\rm sym}) =  \hat Z( \vert \{ p\}_{\rm sym})$.
Taking $\Delta=1$,   the running coupling is given by
\begin{eqnarray}
g(t) &=& \frac{u t^{2-d/2}}{1+\dfrac{u}{6} B\left(1/t\right)},
\end{eqnarray}
where $B(p^2) = B(0\vert p)$.

\subsubsection{$d=2$}
In 2-dimensions, we obtain
\begin{eqnarray}
g(t)&=& \frac{ut}{1+\dfrac{ut}{6\pi\sqrt{1+4m^2t}}\tanh^{-1}\left(\dfrac{1}{\sqrt{1+4m^2t}}\right)},
\end{eqnarray}
which behaves in  the UV limit ($t\rightarrow 0$) and IR limit ($t\rightarrow\infty$) as
\begin{eqnarray}
g(t) &\simeq &\left\{
\begin{array}{ccc}
\dfrac{ ut}{1- u t \log (m^2 t) /(12\pi)}  & \rightarrow 0, & t=0 \\
\\
\dfrac{u t}{1+u/(24\pi m^2)} &\rightarrow \infty , & t=\infty \\
\end{array}
\right.  .
\end{eqnarray}
In the massless limit $m^2\rightarrow 0$, we have
\begin{eqnarray}
g(t) &\simeq &- \frac{12\pi}{\log (m^2 t)}\rightarrow 0.
\end{eqnarray}

\subsubsection{$d=3$}
At $d=3$, the running coupling is given by
\begin{eqnarray}
g(t) &=&\dfrac{ u\sqrt{t}}{1+\dfrac{u\sqrt{t}}{24\pi}\arctan\left(\dfrac{1}{\sqrt{4m^2 t}}\right)},
\end{eqnarray}
which behaves as
\begin{eqnarray}
g(t) &\simeq &\left\{
\begin{array}{ccc}
\dfrac{ u\sqrt{t}}{1+ u \sqrt{t}/48}  & \rightarrow 0, & t=0 \\
\\
\dfrac{u\sqrt{t}}{1+u /(48\pi m)} &\rightarrow \infty , & t=\infty \\
\end{array}
\right.  .
\end{eqnarray}

In the massless limit, we have
\begin{eqnarray}
g(t) &=& \dfrac{ u\sqrt{t}}{1+ u \sqrt{t} /48} =
\left\{
\begin{array}{cc}
\rightarrow 0, & t\rightarrow 0 \\
\rightarrow 48, & t\rightarrow\infty \\
\end{array}
\right.  ,
\end{eqnarray}
which correspond to the asymptotic free UV fixed point and the Wilson-Fisher IR fixed point, respectively.

\subsubsection{$d\ge 4$}
Since $B(Q^2)$ diverges as $\Lambda^{d-4}$ ($\log \Lambda$ at $d=4$) at $d\ge 4$, the running coupling vanishes as the cut-off is removed ($\Lambda\rightarrow\infty$). 
Thus the theory is trivial in the continuum limit at $d\ge 4$.

\subsection{Interacting flow in the massless limit at $d=3$}
\subsubsection{Massless limit}
We next consider the interacting flow case, where we need to evaluate $\hat X$ and $\hat Y$, which are difficult to calculate in general. We therefore consider the massless limit.\footnote{We will indicate the massless limit by a subscript $_0$.}  In this limit,  the kernel function is reduced to
\begin{eqnarray}
K(t,s\vert \{ p \}_{\rm sym.}) &=& D^{d/2-1} k_0(D t, D s),
\end{eqnarray}
where 
\begin{eqnarray}
k_0(w,v) &=& \frac{\rme^{v-w} w^{1-d/2}}{2^{d-1}(2\pi)^{d/2}}\int_0^1  \rmd z\, z^{d/2-2}\exp\left[\dfrac{(w-v)^2 z}{2w}\right],
\end{eqnarray}
and we regard $D\equiv Q^2=\Delta/t$ as an independent  variable. Here the $z$ integral is convergent for $d>2$ while the bubble integral $B(0\vert Q)$ is finite for $d<4$.
We thus concentrate on the $d=3$ case hereafter.

In this limit, we obtain (see appendix~\ref{app:massless} for details)
\begin{eqnarray}
\hat  Z(\vert  \{ p \}_{\rm sym.}) &=& -16\sqrt{D}\frac{\bar u(D) }{1 + \bar u (D)}, \qquad \bar u(D) \equiv \frac{u}{48\sqrt{D}}, \label{eq:hatZ}\\
\hat Y(t\vert \{ p \}_{\rm sym.}) &=& \frac{3 }{4}\sqrt{D}  \left\{\xi_0^{(1)}(\Delta) - 8\xi_0^{(2)}(\Delta)
\frac{\bar u(D)}{1+\bar u(D)}\right\}, \label{eq:hatY}\\
\hat X(t,t\vert  \{ p \}_{\rm sym.}) &=& \frac{9}{16}\sqrt{D}\left\{ \Xi_0(\Delta) -4\{\xi_0^{(2)}(\Delta)\}^2\frac{\bar u(D)}{1+\bar u(D)}\right\}, \label{eq:hatX}
\end{eqnarray}
where 
\begin{eqnarray}
\xi_0^{(i)}(\Delta) &=& \int_0^\Delta  \rmd w\, \phi_0^{(i)}(w), \quad i=1,2, \\
\Xi_0(\Delta) &=&\int_0^\Delta  \rmd w\, \int_0^\Delta  \rmd v\, \Omega_0(w,v) ,
\end{eqnarray}
and $\phi_0^{(i)}$ and $\Omega_0$ are solutions to the integral equations
\begin{eqnarray}
\rme^{-3w/2} + \int_0^w  \rmd v\, k_0(w,v)\, \phi_0^{(1)}(v) &=& 0,  \label{eq:phi0_1}\\
b_0(w) +\int_0^w  \rmd v\, k_0(w,v)\, \phi_0^{(2)}(v) &=& 0 , \label{eq:phi0_2}\\
b_0(w+v) -2 \int_0^w  \rmd x\, k_0(w,x) \int_0^v  \rmd y\, k_0(v,y)\, \Omega_0 (x,y) &=& 0,
\label{eq:Omega0}
\end{eqnarray}
 where $b_0(w)$ is the massless bubble integral given by eq.~(\ref{eq:b0}). 
These equations can be solved numerically, and 
at $\Delta =1$, for example, we have  $\xi_0^{(1)}(1) = -14.8440(1)$,  $\xi_0^{(2)}(1) =  -1.60557(1)$ and
$\Xi_0(1) =  16.6753(1)$.

\subsubsection{Running coupling and $\beta$ function}
Using the above results, the running coupling at $d=3$ is given by
\begin{eqnarray}
g_0(\mu) &=& G_1 + G_2 \frac{\bar u(\Delta) \sqrt{t}}{1+ \bar u(\Delta) \sqrt{t}}, \qquad \bar u(\Delta)= \frac{ u}{48\sqrt{\Delta}} ,
\end{eqnarray}
where $\mu =1/\sqrt{t}$ and
\begin{eqnarray}
G_1&=& -9\sqrt{\Delta} \left[ \xi_0^{(1)}(\Delta) +\frac{3}{4}\Xi_0(\Delta) \right], \quad
G_2 = 48\sqrt{\Delta} \left[ 1 +\frac{3}{4}\xi_0^{(2)}(\Delta)\right]^2  \ge\ 0 . 
\end{eqnarray}
 With the numerical values given above we obtain $G_1 = 21.0378(1)$ and $G_2 = 2.00105(1)$ at $\Delta=1$.
\footnote{It turns out that $G_2(\Delta)$ has only one zero at
  $\Delta=0.36228(1)$.}

We then calculate the $\beta$ function for $ g_0( \mu )$  as 
\begin{eqnarray}
\beta(g_0)&\equiv&\mu \frac{\partial }{\partial \mu} g_0(\mu) = \frac{(g_0(\mu) - G_1-G_2)(g_0(\mu) - G_1)}{G_2}, 
\end{eqnarray}
which becomes zero at $g_0(\mu) =G_1$ and $g_0(\mu) =G_1+G_2$.
The coupling  $g_0(\mu)$ near  $G_1$ behaves as 
\begin{eqnarray}
g_0(\mu) - G_1 \simeq C_{UV} \frac{u}{\mu} \rightarrow 0 , \qquad \mu\rightarrow \infty, \qquad
C_{UV} = \left[ 1 +\frac{3}{4}\xi_0^{(2)}(\Delta)\right]^2,
\end{eqnarray}
approaching to the UV fixed point from above, while  near $G_1+G_2$ we have the IR fixed point as 
\begin{eqnarray}
g_0(\mu) - G_1-G_2 \simeq - C_{IR} \frac{\mu}{u} \rightarrow 0 , \qquad \mu\rightarrow 0, \qquad
C_{IR} =\left\{48\sqrt{\Delta}  \left[ 1 +\frac{3}{4}\xi_0^{(2)}(\Delta)\right]\right\}^2,
\end{eqnarray}
where the coupling approaches from below to  the Wilson-Fisher fixed point in the 3 dimensional scalar theory.
Note that the derivative of the $\beta$ function with respect to $g_0$ at the fixed point becomes
\begin{eqnarray}
\beta^\prime(g_0) \equiv  \frac{d\beta(g_0)}{d g_0} &=& 
\left\{
\begin{array}{ll}
  -1, &  g_0=G_1     \\
  1, & g_0 = G_1+G_2    \\
\end{array}
\right. ,
\end{eqnarray}
which should be compared with the same quantities calculated for the standard running coupling in the 3 dimensional massless theory in Ref.~\cite{Aoki:2014yra}, where  $\beta^\prime (0) = -1$ (UV) and $\beta^\prime (48) = 1$ (IR). 
The derivative of the $\beta$ function at the fixed point gives the anomalous dimension of the operator conjugate to the coupling in the conformal theory at the fixed point, and thus is universal.
Our flow coupling indeed satisfies this condition and  the derivatives at the two fixed points agree with those for the conventional definition of the coupling.
This establishes that  our flow coupling gives a good definition of the running coupling of the theory.
 The scaling dimension $\gamma$ of the operator  conjugate to the running coupling $g_0$ is given by $\gamma=d+\beta^\prime(g_0)$, so that $\gamma_{\rm UV}= 2$ and $\gamma_{IR}=4$ in this model.
Interestingly $\gamma_{\rm UV}= 2$ corresponds to the canonical dimension of the $\varphi^4$ operator in 3 dimensions, which is  the interaction term in our theory.

By the redefinition of the coupling as $g(\mu) \equiv ( g_0(\mu) - G_1 )/G_2$, the corresponding $\beta$ function is simplified as
\begin{eqnarray}
\beta (g) &\equiv&\mu \frac{\partial }{\partial \mu} g(\mu) = g(\mu) (g(\mu) - 1) .
\end{eqnarray}
 
\section{NLO corrections to the induced metric}
\label{sec:metric}

In Ref.~\cite{Aoki:2016ohw}, the induced metric has been calculated from the flowed scalar field in the large $N$ limit.
It has been shown that the metric from the massive scalar field describes a space which becomes  the Euclidean AdS space asymptotically both in UV and IR limits, where the radius $R_{\rm IR}$ in the IR is larger than the radius $R_{\rm UV}$ in UV as
\begin{eqnarray}
R_{\rm UV}^2 &=&\frac{d-2}{2} R_{\rm IR}^2 < R_{\rm IR}^2,
\end{eqnarray}
while the metric describes the whole AdS space in the massless limit with the radius $R_{\rm UV}$.
In this section, we consider the NLO correction to the induced metric in the $1/N$ expansion
as another application of the NLO calculation,
in particular, in the massless case at $d=3$, in order to see whether the space remains AdS or not and how the radius changes at the NLO. 

\subsection{Induced metric at NLO}
The VEV of  the induced metric is defined from the normalized flowed field as\cite{Aoki:2016ohw}
\begin{eqnarray}
g_{\mu\nu} (z) &=& R_0^2 \langle \partial_\mu \sigma^a (z) \partial_\nu \sigma^a (z) \rangle 
\end{eqnarray}
with some length scale $R_0$, where $z=(\tau=2\sqrt{t}, x)$ and $\mu,\nu =0,1,\cdots, d$.
Here $\sigma^a(z)$ is the normalized flowed field such that $\langle \sigma^2(z) \rangle =1$, and the corresponding 2-point function is explicitly given at NLO as
\begin{eqnarray}
\langle \sigma^{a_1}(z_1)  \sigma^{a_2}(z_2) \rangle &=& \frac{\delta^{a_1a_2}}{N}\frac{1}{\sqrt{\zeta_0(t_1)\zeta_0(t_2)}}\left( 1-\dfrac{\zeta_1(t_1)+\zeta_1(t_2)}{2N}\right) \nonumber \\
&\times&
\int \rmD p \frac{\rme^{-p^2(t_1+t_2)}\rme^{i p(x_1-x_2)}}{p^2+m^2} \left[1 +\frac{G_1(t_1,t_2\vert p)}{N} \right],
\end{eqnarray}
where
\begin{eqnarray}
\zeta_1(t) &=& \frac{1}{\zeta_0(t)} {\cal H}\left[ G_1(t,t\vert p) \right] , \qquad  {\cal H}\left[ f(t\vert p) \right]\equiv\int \rmD p \frac{\rme^{-2p^2t} }{p^2+m^2} f(t\vert p) . 
\label{eq:def_H}
\end{eqnarray}

After some  algebra (see appendix~\ref{app:metric}), we obtain
\begin{eqnarray}
g_{ij}(\tau) &=& \delta_{ij}\frac{ R_0^2}{d} A(t), \quad (i,j=1, 2, \cdots, d), \qquad g_{00}(\tau) = - \frac{ R_0^2\,  t}{2}\partial_t A(t), 
\end{eqnarray}
where
\begin{eqnarray}
A(t) &=& -\frac{1}{2} \frac{\partial_t\zeta_0(t)}{\zeta_0(t)} +\frac{1}{N}A_1(t), 
\end{eqnarray}
and $A_1(t)$ in general is a very complicated function given in appendix~\ref{app:metric}.

\subsection{Induced metric in the massless limit at $d=3$}
In the massless limit at $d=3$, the metric at the LO  is  given by 
 \begin{eqnarray}
g_{ij}(\tau) &=&\delta_{ij} \frac{R_0^2}{3\tau^2}, \qquad g_{00}(\tau) = \frac{R_0^2}{2\tau^2},
\end{eqnarray}
which describes the AdS space for all $\tau$.

At the NLO, $A_1(t)$ is given by
\begin{eqnarray}
A_1(t) &=&  \frac{1}{2\sqrt{t}}\int \rmD Q\, h_{\rm total}(Q^2)\frac{\bar u(Q^2)}{(1+\bar u(Q^2)\sqrt{t})^2}, 
\quad \bar u(Q^2) = \frac{u}{48\sqrt{Q^2}}, \label{eq:A1} \\
\partial_t A_1(t) &=&  -\frac{1}{4\sqrt{t^3}}\int \rmD Q\, h_{\rm total}(Q^2)\frac{\bar u(Q^2)(1+3\bar u(Q^2)\sqrt{t})}{(1+\bar u(Q^2)\sqrt{t})^3}, \label{eq:pA1} 
\end{eqnarray}
which leads to
\begin{eqnarray}
g_{ij}(\tau) &=&\delta_{ij}\frac{R_0^2}{3 \tau^2} \left[1 +\frac{\tau}{N}\int \rmD Q\, h_{\rm total}(Q^2)
\frac{\bar u(Q^2)}{(1+\bar u(Q^2)\tau/2)^2}\right], 
\label{eq:metric_space}\\
g_{00}(\tau) &=& \frac{R_0^2}{2\tau^2} \left[1 +\frac{\tau}{2N}\int \rmD Q\, h_{\rm total}(Q^2)
\frac{\bar u(Q^2)(1+3\bar u(Q^2)\tau/2)}{(1+\bar u(Q^2)\tau/2)^3} \right] ,
\label{eq:metric_time}
\end{eqnarray}
where $h_{\rm total}(Q^2)$ is a function given in appendix~\ref{app:metric}. 

\subsection{UV and IR limits}
The above expression  in the UV limit ($\tau\rightarrow 0$) leads to
\begin{eqnarray}
g_{ij}(\tau) &\simeq&\delta_{ij}\frac{R_0^2}{3\tau^2} \left[1 +\frac{\tau}{N}\int \rmD Q\, h_{\rm total}(Q^2)
\bar u(Q^2)\right], \quad \tau\rightarrow 0, \\
g_{00}(\tau) &\simeq& \frac{R_0^2}{2 \tau^2} \left[1 +\frac{\tau}{2N}\int \rmD Q\, h_{\rm total}(Q^2)
\bar u(Q^2) \right] , \quad \tau\rightarrow 0,
\end{eqnarray}
which shows that the NLO correction is less singular than the LO contribution. Therefore
the space becomes asymptotically AdS in the UV limit at the NLO whose AdS radius is equal to that at the LO.  

We cannot naively take the $\tau\rightarrow\infty$ limit in eqs.~(\ref{eq:metric_space}) and (\ref{eq:metric_time}), 
on the other hand,  due to the enhancement of the UV contribution of the $Q$ integrals.
Careful evaluations of these $Q$ integrals in appendix~\ref{app:metric} give
\begin{eqnarray}
g_{ij}(\tau) &\simeq&\delta_{ij}\frac{R_0^2}{3\tau^2} \left[1 +\frac{r}{N}\right], \quad 
g_{00}(\tau) \simeq \frac{R_0^2}{2 \tau^2} \left[1 +\frac{r}{N}\right], \quad \tau\rightarrow \infty, 
\end{eqnarray}
where $r= -0.41869(1)$.\footnote{This is independent of  $u_f\not= 0$ (the interacting flow).
In the case of free flow ($u_f=0$), however, $r=\dfrac{8}{3\pi^2}\simeq0.27019$.}  
 Therefore, the space  becomes asymptotically AdS again in the IR limit, whose radius, however, is smaller than that in the UV limit.\footnote{It is interesting and also suggestive to see that the F-coefficient of the 3 dimensional $O(N)$ scalar model is given by $F_{\rm IR} = F_{UV} - \zeta(3)/(8\pi^2) + O(1/N)$, where $F_{\rm UV} = N F_S$ with $F_S\simeq 0.0638$ as an example of a conjecture, the so-called "the F-theorem", which claims that the F-coefficient monotonically decreases along a RG trajectory connecting two 3 dimensional  CFTs. Furthermore, in the holographic dual picture,  the F-coefficient is proportional to the AdS radius squared. (See Ref.~\cite{Pufu:2016zxm} and references therein.)
 }
The induced metric at the NLO describes a 4 dimensional space which is asymptotically AdS in both UV and IR regions with different radii but non-AdS in-between.

It is clear that the NLO correction to the AdS radius in the IR limit is related to the Wilson-Fisher fixed point in the original 3 dimensional scalar theory, since the eqs.~(\ref{eq:metric_space}) and (\ref{eq:metric_time}) can be written as
\begin{eqnarray}
g_{ij}(\tau) &=& \delta_{ij}\frac{R_0^2}{3 \tau^2} \left[1 -\frac{1}{24N}\int \rmD Q\, h_{\rm total}(Q^2)
\beta ( g ( 48 \mu\sqrt{Q^2}))\right], \\
g_{00}(\tau) &=& \frac{R_0^2}{2\tau^2} \left[1 -\frac{1}{24N}\int \rmD Q\, h_{\rm total}(Q^2)
\left\{ 1 +  \frac{\mu}{2}\frac{ \partial}{\partial \mu}\right\} \beta ( g (  48\mu\sqrt{Q^2}))
 \right] ,
\end{eqnarray}
where $\mu =1/\sqrt{t} =2/\tau$, and $\beta(g(x))$ is the $\beta$ function for the running coupling $g(x)$ from the free flowed field defined in the previous section with $\Delta=1$ as
\begin{eqnarray}
\beta(g) &=& \frac{g(g-48)}{48}, \qquad 
g(x) = 48 \frac{ u }{x+ u }.
\end{eqnarray}

\section{Summary}
\label{sec:summary}
In this paper, 
we studied the flow equation of the O($N$) $\varphi^4$ model in $d$ dimensions at the NLO in the $1/N$ expansion,
employing the Schwinger-Dyson equation.
We calculated the 2-pt and 4-pt functions at the NLO.

As an application of the NLO calculation,
 we investigated the running coupling defined from the connected 4-pt function of flowed fields.
In particular at $d=3$ in the massless limit, we showed that the running coupling has two fixed points, the asymptotic free one in the UV region and the Wilson-Fisher one in the IR region.
We also derived the corresponding $\beta$ function.
Our study suggests that the flow equation can be interpreted as a renormalization group transformation.

We also calculated the NLO correction to the $d+1$ dimensional metric induced  from the massless scalar field theory at $d=3$.
In the massless limit, the whole 4-dimensional space becomes AdS at the LO of the  $1/N$ expansion\cite{Aoki:2016ohw}.
We found that the NLO corrections give small perturbations to the metric,
which make the space only asymptotically AdS in both UV ($t=0$) and IR ($t=\infty$) limits.
In addition,  while the NLO corrections do not change the AdS radius at the LO in the UV limit, the AdS radius is  reduced by the NLO correction in the IR limit, which corresponds to the Wilson-Fisher IR fixed point of the original theory.  The nontrivial fixed point in the field theory appears as a change of the AdS radius at the NLO.
The induced metric at NLO describes a 4-dimensional space which connects one asymptotically AdS space at UV to the other  asymptotically AdS space at IR.

This paper contains two important messages.
One is that the flow equation can provide an alternative method  to define a renormalization group transformation.
The other is that the massless scalar field in $d$ dimensions plus  the extra dimension from  the RG scale
not only generates  a $d+1$ dimensional AdS space at  LO\cite{Aoki:2016ohw} but also 
gives a NLO correction, which makes the $d+1$ dimensional space  asymptotically AdS only in UV and IR limits at $d=3$.   In particular, the AdS radius in the IR limit, which corresponds to the Wilson-Fisher fixed point, becomes smaller than that in the UV limit, which is equal to the radius at the LO.
Although the relation found in this paper between the massless scalar field theory and the induced geometry is very suggestive, further studies will be needed to establish an alternative interpretation  of  AdS/CFT correspondences
proposed in Ref.~\cite{Aoki:2015dla} in terms of field theories.

\section*{Acknowledgement}
The authors  would like to thank Satoshi Yamaguchi for very useful comments and discussions.
S. A. is supported in part by the Grant-in-Aid of the Japanese Ministry of Education, Sciences and Technology, Sports and Culture (MEXT) for Scientific Research (No. JP16H03978),  
by a priority issue (Elucidation of the fundamental laws and evolution of the universe) to be tackled by using Post ``K" Computer, 
and by Joint Institute for Computational Fundamental Science (JICFuS).
This investigation has also been supported in part by the Hungarian 
National Science Fund OTKA (under K116505).
S. A. and J. B. would like to thank the Max-Planck-Institut f\"ur Physik 
for its kind hospitality during their stay for this research project.
T.O. is supported in part by the Grant-in-Aid of the Japanese 
Ministry of Education, Sciences and Technology, 
Sports and Culture (MEXT) for Scientific Research (No. 26400248).


%

\appendix

\section{The $1/N$ expansion in the $d$ dimensional theory}   
\label{app:d-dim}
In this appendix, we consider the $1/N$ expansion in the $d$ dimensional theory.

\subsection{Schwinger-Dyson equation(SDE)}
In order to perform the $1/N$ expansion, we consider the   SDE of this model, which can be written compactly as
\begin{eqnarray}
\langle \delta^a_x X[\varphi] \rangle &=& \langle  X[\varphi] \delta^a_x S(\mu^2,u) \rangle,
\end{eqnarray}
where $ \delta^a_x \varphi^b(y) = \delta^{ab}\delta^{(d)}(x-y) \epsilon$
with a small parameter $\epsilon$, so that
\begin{eqnarray}
\delta^a_x S(\mu^2,u) &=& N \epsilon\left[ (- \Box +\mu^2)\varphi^a(x) + \frac{u}{3!}\varphi^a(x)\varphi^2(x)\right] .
\end{eqnarray}
Here  the vacuum expectation value of an operator ${\cal O}$  is defined in eq.~(\ref{eq:vev_d}).

We define $2n$-point  functions $\Gamma_{2n}$\footnote{Note that we use the same notation $\Gamma_{2n}$ for the 2$n$-point functions in both $d$ and $d+1$ dimensions, since no confusion may occur.}
as
\begin{eqnarray}
\Gamma^{a_1a_2\cdots a_{2n}}(x_1,x_2,\cdots,x_{2n}) &=&
N^{2n-1}\left\langle \prod_{i=1}^{2n} \varphi^{a_i}(x_i) \right\rangle \equiv 
\Gamma_{2n}[12\cdots (2n) ]
\label{eq:G2n_d}
\end{eqnarray}
which can be written in  terms of their connected parts $K_{2n}$ as
\begin{eqnarray}
\Gamma_4[1234] &=& K_4[1234] + N\left\{\Gamma_2[12]\Gamma_2[34] + \Gamma_2[13]\Gamma_2[24]
+\Gamma_2[14]\Gamma_2[23]   \right\}, \\
\Gamma_6[123456] &=& K_6[123456] + N\left\{\Gamma_2[12]K_4[3456] + \mbox{ 14 perms. } \right\}\nonumber \\
&+& N^2\left\{\Gamma_2[12] \Gamma_2[34]\Gamma_2[56]+ \mbox{ 14 perms. } \right\}
\end{eqnarray}
and so on. 
As mentioned in the main text, 
we assume we are working in a phase where O$(N)$ symmetry is not broken.
We therefore do not add the external source term $h \varphi(x)$ to the action, 
so that the action has the symmetry under $\varphi \rightarrow -\varphi$, which implies
$\Gamma_{2n-1} = 0$ for all positive integers $n$. 

In terms of these, the SDE for $X(\varphi) = \varphi^{a_2}(x_2)$ becomes
\begin{eqnarray}
\delta_{12} &=&  (- \Box +\mu^2)_{x_1} \Gamma_2[12] +\frac{u}{3! N^2}\sum_b \left(K_4[bb12]
+N\left\{ \Gamma_2[bb]\Gamma_2[12] + 2\Gamma_2[b1]\Gamma_2[b2]\right\} \right)~~~
\label{eq:SDE_2pt}
\end{eqnarray}
where $\delta_{12}\equiv\delta^{a_1a_2}\delta^{(d)}(x_1-x_2)$ and $x_b=x_1$, so that $b$ in the summation runs over the O($N$) indices only.

For $X(\varphi) = \varphi^{a_2}(x_2) \varphi^{a_3}(x_3)\varphi^{a_4}(x_4)$, on the other hand, we have
\begin{eqnarray}
\delta_{12}\Gamma_2[34] +
\mbox{ 2 perms.}
 &=& (- \Box +\mu^2)_{x_1} \frac{1}{N}\left( K_4[1234]+ N\left\{\Gamma_2[12]\Gamma_2[34] + \mbox{ 2 perms.}\right\}\right) \nonumber \\
&+& 
\frac{u}{3! N^3}\sum_b \left(K_6[bb1234] + N\left\{\Gamma_2[bb]K_4[1234] + \mbox{ 14 perms.} \right\}\right.
\nonumber \\
&+&\left. N^2\left\{\Gamma_2[bb]\Gamma_2[12]\Gamma_2[34]+\mbox{ 14 perms.} \right\}\right),
\end{eqnarray}
which can be simplified by using eq.~(\ref{eq:SDE_2pt}) as
\begin{eqnarray}
0&=&  (- \Box +\mu^2)_{x_1}  K_4[1234] + \frac{u}{3! N^2}\sum_b \Bigl( K_6[bb1234]
+ N\Gamma_2[bb]K_4[1234] \nonumber \\
&+& 2N\left\{\Gamma_2[b1]K_4[b234] + \Gamma_2[b2]K_4[1b34]+\Gamma_2[b3]K_4[12b4]+\Gamma_2[b4]K_4[123b]
\right\}  \nonumber \\
&+&N\left\{ \Gamma_2[12]K_4[bb34] + \Gamma_2[13]K_4[b2b4]+\Gamma_2[14]K_4[b23b]
\right\} \nonumber \\  
&+& 2N^2\left\{\Gamma_2[b2][\Gamma_2[b3]\Gamma_2[14] + \Gamma_2[b2][\Gamma_2[b4]\Gamma_2[13]
+\Gamma_2[b3][\Gamma_2[b4]\Gamma_2[12]  
\right\}  \Bigr)  .
\label{eq:SDE_4pt}
\end{eqnarray}

Using the O($N$) symmetry and assuming translational invariance (e.g. infinite volume or periodic boundary condition), we can write
\begin{eqnarray}
\Gamma_2[12]&\equiv& \delta^{a_1a_2}\Gamma(x_{12}),\qquad x_{12}\equiv x_1-x_2 \\
K_4[1234]&\equiv& \delta^{a_1a_2} \delta^{a_3a_4} K(x_1,x_2;x_3,x_4) + \mbox{ 2 perms.} , \\
K_6[123456] &\equiv& \delta^{a_1a_2} \delta^{a_3a_4}  \delta^{a_5a_6} H(x_1,x_2;x_3,x_4;x_5,x_6) 
+  \mbox{ 14 perms.},
\end{eqnarray}
where $K(x_1,x_2;x_3,x_4)$ is invariant under $1\leftrightarrow 2$ or $3\leftrightarrow 4$ as well as
$(12)\leftrightarrow (34)$, and similar invariances hold for $H(x_1,x_2;x_3,x_4;x_5,x_6)$.

We finally obtain
\begin{eqnarray}
\delta^{(d)}(x_1-x_2) &=&  \left[ (- \Box +\mu^2)_{x_1} +\frac{u}{3!} \Gamma(0)\right] \Gamma(x_{12})\nonumber \\
&+& \frac{u}{3! N}\left[\left(1+\frac{2}{N}\right) K(x_1,x_1;x_1,x_2) +  2\Gamma(0)\Gamma(x_{12})\right] ,
\label{eq:SDE_2ptA}
\end{eqnarray}
and
\begin{eqnarray}
0&=&  \left[ (- \Box +\mu^2)_{x_1} +\frac{u}{3!}\left(1+\frac{2}{N}\right)\Gamma(0)\right] K(x_1,x_2;x_3,x_4)
\nonumber \\
&+& 
\frac{ u}{3! }\Gamma(x_{12}) \left[ 
2 \Gamma(x_{13})\Gamma(x_{14}) +\left(1+\frac{2}{N}\right) K(x_1,x_1;x_3,x_4)
+\frac{2}{N}K(x_1,x_3;x_1,x_4)\right]
\nonumber \\
&+& \frac{u}{3! N}\left[\left(1+\frac{2}{N}\right) H(x_1,x_1;x_1,x_2;x_3,x_4) +  \frac{2}{N}H(x_1,x_2;x_1,x_3;x_1,x_4)\right]\nonumber\\
&+&  \frac{2u}{3! N}
\left[  \Gamma(x_{13})K(x_1,x_2;x_1,x_4)+ \Gamma(x_{14})K(x_1,x_2;x_3,x_1) \right] .~~~~~
\label{eq:SDE_4ptA}
\end{eqnarray}

\subsection{The leading order in the $1/N$ expansion}
We introduce the $1/N$ expansion as
\begin{eqnarray}
\Gamma(x_{12}) &=&\sum_{i=0}^\infty N^{-i} \Gamma_i(x_{12}), \qquad
K(x_1,x_2;x_3,x_4) = \sum_{i=0}^\infty N^{-i} K_i(x_1,x_2;x_3,x_4), 
\end{eqnarray}
and so on, together with $\displaystyle \mu^2 =\sum_{i=0}^\infty N^{-i} \mu^2_i $.

At  the leading order (LO) of the $1/N$ expansion,
the eq.~(\ref{eq:SDE_2ptA}) in momentum space becomes  
\begin{eqnarray} 
1 &=& \left( p^2+\mu^2_0 +\frac{u}{6}  \int \rmD q\,  \widetilde{\Gamma}_0(q) \right)   \widetilde{\Gamma}_0(p) ,
\qquad   \Gamma_0(x) =\int Dp\, \widetilde{\Gamma}_0(p)\, \rme^{ipx}, 
\end{eqnarray}
which can easily be solved as
\begin{eqnarray} 
  \widetilde{\Gamma}_0(p) &=&\frac{1}{p^2+m^2}, \qquad m^2 = \mu^2_0 + \frac{u}{6} Z(m), 
 \label{eq:sol_2pt}
\end{eqnarray}
where $m \ge 0$ is the renormalized mass  and $Z(m)$ is given in eq.~(\ref{eq:Zm}). 
Thus the 2-pt function at the LO becomes
\begin{eqnarray}
\langle \varphi^a(x) \varphi^b(y)\rangle &=& \frac{\delta^{ab}}{N}\int \rmD p \frac{\rme^{ip(x-y)}}{p^2+m^2} .
\end{eqnarray}

 Eq.~(\ref{eq:SDE_4ptA}) at the LO  leads to
\begin{eqnarray}
 (- \Box +m^2)_{x_1} K_0(x_1,x_2;x_3,x_4)
&+&\frac{ u}{3! } \Gamma_0(x_{12})K_0(x_1,x_1;x_3,x_4)\nonumber \\
&=& -\frac{ 2 u}{3! } 
\Gamma_0(x_{12})\Gamma_0(x_{13})\Gamma_0(x_{14}) .
\end{eqnarray}
 Introducing a function $G_0(p_1,p_2,p_3,p_4)$ to rewrite $K_0(x_1,x_2,x_3,x_4)$ as
\begin{eqnarray}
K_0(x_1,x_2;x_3,x_4) &=& \left\{\prod_{i=1}^4 \int  \rmD p_i \frac{\rme^{i p_i x_i}}{p_i^2+m^2}\right\}
G_0(p_1,p_2,p_3,p_4)(2\pi)^d \delta(p_1+p_2+p_3+p_4),~~~~~~~~
\end{eqnarray}
we obtain 
\begin{eqnarray} 
G_0(p_1,p_2,p_3,p_4) &=& G_0(p_1+p_2) = -\frac{2 u}{6+ u B(p_{12}^2)},
\label{eq:4pt_LO}
\end{eqnarray}
where $p_{12}=p_1+p_2$, and 
\begin{eqnarray}
B(Q^2) &=& \int {\rmD}q_1 {\rmD}q_2 \frac{(2\pi)^d\delta(q_1+q_2-Q)}{(q_1^2+m^2)(q_2^2+m^2)} 
= \int_0^1  \rmd x\,  \int {\rmD}q_1 \frac{ \theta(\Lambda^2-q_1^2)}{(q_1^2+m^2+ x(1-x) Q^2)^2}  . ~~~~~~~
\end{eqnarray}
 This agrees with the previous result obtained by a different method\cite{Aoki:2014yra}.
We here specify the way we introduce the cut-off $\Lambda$ for the case where $B(Q^2)$ diverges.

\subsection{NLO correction to the 2-pt functions}
Let us consider the next-to-leading order (NLO) correction to the 2-pt function $\Gamma_2$.
At the NLO, eq.~(\ref{eq:SDE_2ptA}) leads to
\begin{eqnarray}
0&=& (-\Box +m^2)\Gamma_1(x_{12}) +\left\{\frac{u}{6}(2 Z(m) + \gamma_1) +\mu^2_1\right\} \Gamma_0(x_{12}) +\frac{u}{6}K_0(x_1,x_1;x_1,x_2), ~~~~ \\
\gamma_1 &=& \int \rmD q\, \widetilde{\Gamma}_1(q), 
\end{eqnarray}
which can be solved in  momentum space as
\begin{eqnarray}
\widetilde{\Gamma}_1(p) &=& -\frac{1}{(p^2+m^2)^2}\left( \mu^2_1+\frac{u}{6}\gamma_1 + \frac{u}{3} S(p^2)\right), 
\label{eq:Gamma_1}
\end{eqnarray}
where
\begin{eqnarray}
S(p^2) =\int \frac{ \rmD Q }{(p-Q)^2+m^2}\frac{6}{6+u B(Q^2)} ,
\end{eqnarray}
and the condition for $\gamma_1$ is solved as
\begin{eqnarray}
\gamma_1 &=& - \frac{\mu_1^2 B(0) +C_2}{1+\frac{u}{6} B(0)}, \qquad
C_2\equiv - \int \frac{ \rmD Q }{\frac{6}{u}+ B(Q^2)}\frac{d }{d m^2} B(Q^2) .
\label{eq:gamma_1}
\end{eqnarray}
 Substituting  eq.~(\ref{eq:gamma_1}) into eq.~(\ref{eq:Gamma_1}), we finally obtain 
\begin{eqnarray}
\widetilde{\Gamma}_1(p) &=& - \frac{1}{(p^2+m^2)^2}\left\{ g(p^2) + \tilde C \right\},
\end{eqnarray}
where
\begin{eqnarray}
g(p^2) &=& \int \frac{\rmD Q}{\frac{6}{u} +B(Q^2)}\left\{ \frac{1}{(Q+p)^2+m^2} +  \frac{1}{(Q-p)^2+m^2} -  \frac{2}{Q^2+m^2}  \right\} , \\
\tilde C &=&  C_1 +\frac{\mu_1^2}{1+\frac{u}{6}B(0)}
-\frac{C_2}{\frac{6}{u}+B(0)} , \quad
C_1 =  \int \frac{\rmD Q}{\frac{6}{u} +B(Q^2)}  \frac{2}{(Q^2+m^2)}, 
\end{eqnarray}
and $g(p^2)$ can be expanded as
\begin{eqnarray}
g(p^2) &=& Z_1 p^2 + \tilde g(p^2), \quad \tilde g(p^2) = O(p^4),
\end{eqnarray}
where
\begin{eqnarray}
Z_1 &=&\frac{2}{d} \int \frac{\rmD Q}{6/u + B(Q^2)}\left[\frac{4-d}{(Q^2+m^2)^2} -\frac{4 m^2}{ (Q^2+m^2)^3}
\right] .
\label{eq:Z1}
\end{eqnarray}

\subsection{Renormalization}
 Let us now consider the renormalization of the theory.

Our renormalization condition for the renormalized 2-pt function $\Gamma_R$  is given in  momentum space as
\begin{eqnarray}
 \widetilde{\Gamma}_R^{-1}(p) &\simeq & p^2+ m^2, \qquad p^2\simeq 0,
\end{eqnarray}
where $m$ is interpreted as the renormalized mass, which is independent  of  both $N$ and the cut-off.
Relating the bare field to the renormalized field by the renormalization constant $Z_R$ as
$Z_R^{1/2} \varphi_R = \varphi$, we explicitly obtain 
\begin{eqnarray}
Z_R \widetilde{\Gamma}_R(p) &=& \widetilde{\Gamma}(p)  =  \frac{1}{p^2+m^2 +\dfrac{1}{N}\Sigma_1(p^2)} + O\left(\frac{1}{N^2}\right) ,
\end{eqnarray}
where
\begin{eqnarray}
\Sigma_1(p^2) &=& Z_1 p^2 + \tilde C +\tilde g(p^2) .
\end{eqnarray}

At the LO of the $1/N$ expansion, the above condition implies
\begin{eqnarray}
\mu_0^2 &=& m^2 -\frac{u}{6} Z(m) , \qquad Z_R = 1, 
\end{eqnarray}
where $Z(m)$ 
is potentially divergent at $d > 1$. We therefore introduce the momentum cut-off $\Lambda$ to regulate the integral, and 
$\mu^2_0$ is tuned to cancel
the effect of $Z(m)$ including such divergences, in order to  keep the renormalized mass $m$ finite and constant.
The lattice regularization or dimensional regularization is more consistent than the momentum cut-off, but calculations become much more complicated in the lattice regularization or power divergences are difficult to deal with  in the dimensional regularization. Since the momentum cut-off is enough to see the leading divergences, we adopt it in this paper. 

At the NLO, the renormalization condition implies 
\begin{eqnarray}
Z_R &=& 1 -\frac{Z_1}{N}, \qquad 
\mu_1^2 = \left(1+\frac{u}{6}B(0) \right) Z_1 m^2 + \frac{u}{6} C  ,
\label{eq:Z1_mu1}
\end{eqnarray}
where
\begin{eqnarray}
C
&=&
-\int \frac{\rmD Q}{\frac{6}{u}+B(Q^2)}\left[\frac{d B(Q^2)}{d m^2} +2\frac{\frac{6}{u} + B(0)}{Q^2+m^2}\right].
\label{eq:C}
\end{eqnarray}

The renormalization condition for the coupling, which first appears at the NLO of the $1/N$ expansion, is given by  $G_0(Q^2=s) = -u_r(s)/3$, so that
$u_r(s)$ is regarded as the renormalized  coupling at the scale $s$. 
Eq.~(\ref{eq:4pt_LO}) thus leads to
\begin{eqnarray}
u_r(s) &=& \frac{u}{1+\frac{u}{6} B(s)},
\end{eqnarray}
where 
$B(Q^2)$ is divergent at $d \ge 4$.
Therefore the renormalized  coupling goes to zero as
\begin{eqnarray}
u_r(s) \simeq \frac{6}{B(s)} \rightarrow 0, \quad \Lambda \rightarrow \infty 
\end{eqnarray}
at $d\ge 4$. This  indicates the triviality of the $\varphi^4$ theory at $d\ge 4$. 

\subsection{Renormalization constants}
We here explicitly evaluate the renormalization constants.
\subsubsection{$d=1$}
At $d=1$, $\mu_0^2$ is finite as
\begin{eqnarray}
Z(m) &=&  \frac{1}{\pi m} \arctan \left(\frac{\Lambda}{m} \right) 
\end{eqnarray} 
is finite, and
the  coupling is also  finite and nonzero  since
\begin{eqnarray}
B(Q^2) &=&  \frac{1}{ m (Q^2+4m^2)} \simeq \frac{1}{m Q^2}+\cdots,\qquad  Q^2\to\infty,
\end{eqnarray} 
 has a finite limit as $\Lambda\to\infty$.

The most divergent part of $Z_1$ is given by
\begin{eqnarray}
Z_1 &\simeq&\left\{
\begin{array}{ll}
\displaystyle  \int \rmD Q \frac{u}{(Q^2+m^2)^2}, &u\not=\infty \\
\\
\displaystyle  \int \frac{\rmD Q}{B(Q^2)}\frac{6}{(Q^2+m^2)^2} , &u=\infty \\
\end{array}
\right. ,
\end{eqnarray}
which shows that $Z_1$ is finite for all $u$ including $u=\infty$.
 Eqs.~(\ref{eq:Z1_mu1}) and (\ref{eq:C}) thus tell us
that $\mu_1^2$ is also finite for all $u$ including $u=\infty$, and therefore,  there is no divergence at $d=1$ up to the NLO.

\subsubsection{$d=2$}
At $d=2$, 
$\mu_0^2$ is logarithmically divergent as
\begin{eqnarray}  
\mu_0^2 &= & m^2-\frac{u}{6} Z(m), \qquad Z(m)\simeq\frac{1}{ 4\pi} \log \left(\frac{\Lambda^2+m^2}{m^2} \right).
\end{eqnarray} 

On the other hand, $B(Q^2)$ is finite as 
\begin{eqnarray}  
B(Q^2) &=&  \frac{\tanh^{-1}\left(\sqrt{\frac{Q^2}{Q^2+4m^2}}\right)}{\pi \sqrt{Q^2(Q^2+4m^2)}} 
\simeq\frac{1}{2\pi Q^2}\log\frac{Q^2}{m^2} -\frac{m^2}{\pi (Q^2)^2}\left(\log \frac{Q^2}{m^2}-1\right)+\cdots,~~
 \\
\frac{d B(Q^2)}{d m^2} &\simeq& -\frac{2 B(0)}{Q^2 +4m^2}\left[1+\frac{2m^2}{Q^2} \log\frac{Q^2}{m^2}+\cdots\right],
\quad B(0) =\frac{1}{4\pi m^2},
\end{eqnarray} 
so that the renormalized coupling becomes 
\begin{eqnarray} 
u_r(s) &=& \frac{6u}{\displaystyle 6 +u \frac{\tanh^{-1}\left(\sqrt{\frac{s}{s+4m^2}}\right)}{\pi \sqrt{s(s+4m^2)}} } \simeq \frac{12\pi u s}{12\pi s +u \log (s/m^2)}, \qquad s\rightarrow\infty .
\end{eqnarray} 

The most singular term of $Z_1$ for $u\not=\infty$ becomes
\begin{eqnarray}
Z_1 &\simeq&\frac{u}{6} \int {\rmD} Q \frac{2}{(Q^2+m^2)^2},
\end{eqnarray}
which is manifestly finite, while at $u=\infty$, we have
\begin{eqnarray}
Z_1 &=& \int \frac{\rmD Q}{B(Q^2)}\left[\frac{2}{(Q^2+m^2)^2} -\frac{4 m^2}{ (Q^2+m^2)^3}\right],
\end{eqnarray} 
which  diverges  as $Z_1 \simeq  \log\left(\log \Lambda^2\right)$.

The most divergent part of $\mu_1^2$ is given by
\begin{eqnarray}
\mu^2_1 &\simeq&  \left\{
\begin{array}{ll}
-\dfrac{u}{3} Z(m)\delta_1, \quad  (\delta_1=1),  & u \not= \infty \\
\\
 \dfrac{u}{12\pi}\log \left(\dfrac{\Lambda^2+4m^2}{4m^2}\right) , & u\to\infty \\
\end{array}
\right. .
\end{eqnarray}

\subsubsection{$d=3$}
At $d=3$, $\mu_0^2$ is linearly divergent as
\begin{eqnarray}
\mu_0^2 &= & m^2-\frac{u}{6} Z(m), \qquad Z(m)\simeq\frac{1}{ 2\pi^2}\left[\Lambda - m  \arctan \left(\frac{\Lambda}{m} \right) \right],
\end{eqnarray} 
while $B(Q^2)$ is finite as 
\begin{eqnarray}  
B(Q^2) &=&  \frac{1}{4\pi\sqrt{Q^2} } \arctan \left(\sqrt{\frac{Q^2}{4 m^2}}\right)
\simeq\ \frac{1}{8\vert Q\vert} -\frac{ m }{2\pi Q^2} +\frac{2  m^3}{3\pi (Q^2)^2}+\cdots, ~~
 \\
\frac{d B(Q^2)}{d m^2} &=& -\frac{2 B(0)}{Q^2+4m^2},
\quad B(0) =\frac{1}{8\pi m},
\end{eqnarray} 
and the renormalized coupling becomes
\begin{eqnarray} 
u_r(s) &=& \frac{6u}{\displaystyle 6 + \frac{u}{4\pi\sqrt{s} } \arctan \left(\sqrt{\frac{s}{4 m^2}}\right)}
\simeq \frac{ u }{ \displaystyle 1 +\frac{u}{48 \sqrt{s}} }, \qquad s\rightarrow\infty .
\end{eqnarray}

The most singular term of $Z_1$ for $u\not=\infty$ becomes
\begin{eqnarray}
Z_1 &\simeq&\frac{u}{ 9} \int \rmD Q\, \frac{1}{(Q^2+m^2)^2},
\label{eq:Z_1}
\end{eqnarray}
which is manifestly finite at $d=3$. On the other hand, at $u=\infty$, we have
\begin{eqnarray}
Z_1 &=&\frac{2}{3} \int \frac{\rmD Q}{B(Q^2)}\left[\frac{1}{(Q^2+m^2)^2} -\frac{4 m^2}{ (Q^2+m^2)^3}\right],
\end{eqnarray} 
whose divergent part becomes
\begin{eqnarray}
Z_1 &\simeq & 
\dfrac{4}{3\pi^2}\log \Lambda^2 .
\end{eqnarray} 

The most divergent part of $\mu_1^2$ becomes
\begin{eqnarray}
\mu^2_1 &\simeq&\left\{
\begin{array}{ll}
\displaystyle -\frac{u}{3}Z(m)\delta_1, \quad (\delta_1=1), & u\not=\infty \\
\\
 \displaystyle - m\frac{2 u }{9\pi^3}   \log \Lambda^2 ,  & u\rightarrow\infty
\end{array} \right. .
\end{eqnarray}

\subsubsection{$d=4$}
At $d=4$, $\mu_0^2$ is quadratically divergent as
\begin{eqnarray}
\mu_0^2 &= & m^2-\frac{u}{6} Z(m), \qquad Z(m)\simeq \frac{1}{ 16\pi^2}\left[\Lambda^2- m^2\log \left(\frac{\Lambda^2+m^2}{m^2} \right)\right].
\end{eqnarray} 

On the other hand,  at $d=4$, we have
\begin{eqnarray}
B(Q^2)&=&  \frac{1}{(4\pi)^2}\left[ \log\left(\frac{\Lambda_m^2}{m^2}\right)
+2\frac{Q^2+4\Lambda_m^2-2\Lambda^2}{\sqrt{Q^2(Q^2+4\Lambda_m^2)}}\tanh^{-1}
\sqrt{\frac{Q^2}{Q^2+4\Lambda_m^2}}\right. \nonumber \\
&& -\left. 2\sqrt{\frac{Q^2+4m^2}{Q^2}}\tanh^{-1}\sqrt{\frac{Q^2}{Q^2+4m^2}}\right] , \\
B(0) &=& \frac{1}{(4\pi)^2}\left[\log \frac{\Lambda_m^2}{m^2}-\frac{\Lambda^2}{\Lambda_m^2}\right],
\qquad \Lambda_m^2\equiv\Lambda^2+m^2, 
\end{eqnarray}
which diverge logarithmically, so that $u_r(s) =0$ as $\Lambda\rightarrow\infty$.

Since $ \tanh^{-1} (x) \simeq_{x\to 1} -\frac{1}{2}\log\left(\frac{1-x}{2}\right)$,
we have
\begin{eqnarray}
B(Q^2) +\frac{6}{u}&=&\hat B\left(q^2, \alpha^2\right), \quad q^2=\frac{Q^2}{\Lambda^2}, \quad \alpha =\frac{m}{\Lambda}, \\
\hat B(q^2, 0) &=& -c_0 \log q^2 + \frac{6}{u}+ c_0 F(q^2), \quad c_0 = \frac{1}{(4\pi)^2},
\end{eqnarray}
where
\begin{eqnarray}
F(q^2) &=& \frac{2(q^2+2)}{\sqrt{q^2(q^2+4)}}\tanh^{-1}\sqrt{\frac{q^2}{q^2+4}} .
\end{eqnarray}

Let us now consider the continuum limit of $Z_1$. 
By rescaling the momentum, we have
\begin{eqnarray}
Z_1 &= &  -\frac{\alpha^2}{8\pi^2}\int_0^1 \frac{t \rmd t}{\hat B(t,\alpha^2) (t+\alpha^2)^3} .
\label{eq:Z1_4d}
\end{eqnarray}  
As $\alpha^2\rightarrow 0$ in the $\Lambda\rightarrow\infty$ limit, we have 
\begin{eqnarray}
\int_0^1 \frac{ t \rmd t}{\hat B(t,\alpha^2) (t+\alpha^2)^3} &\simeq& \int_0^1 \frac{ t \rmd t}{\hat B(t,0) (t+\alpha^2)^3}\nonumber \\
&=&\int_0^{\frac12} \frac{ t \rmd t}{\hat B(t,0) (t+\alpha^2)^3} + \int_{\frac12}^{1} \frac{ t \rmd t}{\hat B(t,0) (t+\alpha^2)^3},
\end{eqnarray}
where the second term is finite in this limit, while the first term is bounded  from above
\begin{eqnarray}
\int_0^{\frac12} \frac{ t \rmd t}{\hat B(t,0) (t+\alpha^2)^3} &\le &
-\frac{1}{c_0} \int_0^{\frac12}\frac{t \rmd t}{(t+\alpha^2)^3 \log (t+\alpha^2)} \nonumber \\
&=&
\frac{1}{c_0} \left[ \log \vert \log \alpha^2 \vert +\sum_{r=1}^\infty \frac{ (-\log\alpha^2)^r}{r\, r!}
+\mbox{(finite terms)} \right],
\end{eqnarray}
so that $Z_1$ in eq.~(\ref{eq:Z1_4d}) vanishes as $\alpha^2\rightarrow 0$.

The most divergent part of $\mu_1^2$ becomes 
\begin{eqnarray}
\mu^2_1 &\simeq& -\frac{u}{3} \frac{\Lambda^2}{16\pi^2} \delta_1 , \qquad \delta_1 = \int_0^1 { \rmd q^2}\left(\frac{c_0 T(q^2)-6/u}{c_0\{\log q^2 -F(q^2) \} - 6/u}\right),\\
   T(q^2)&\equiv&  \log q^2 +1 -\frac{q^2}{q^2+4}\left( 1+\frac{q^2+6}{q^2+2} F(q^2) \right) ,  \nonumber
\end{eqnarray}
where $\delta_1$ is finite,  but is not universal as it depends on how we regulate the integral.

\subsubsection{$d > 4$}
At $d> 4$, $\mu_0^2$ is $O(\Lambda^{d-2})$ as
\begin{eqnarray}
\mu_0^2 &=& m^2-\frac{u}{6} Z(m), \qquad Z(m)\simeq 
 \frac{d}{ (4\pi)^{d/2}(d-2) \Gamma(1+d/2)}\Lambda^{d-2} .
\end{eqnarray} 

We also write
\begin{eqnarray}
B(Q^2) &=& \frac{d}{(4\pi)^{d/2}\Gamma(1+d/2)}\int_0^1  \rmd x\,\int_0^\Lambda \frac{ p^{d-1} \rmd p}{\left[p^2+m^2 + Q^2 x(1-x)\right]^2} ,
\end{eqnarray}
from which we obtain
\begin{eqnarray}
B(Q^2)&=&  \Lambda^{d-4} \hat B\left(\frac{Q^2}{\Lambda^2}, \frac{m^2}{\Lambda^2}\right), \qquad
\hat B(0,0) =\frac{d}{(d-4)} \frac{1}{(4\pi)^{d/2}\Gamma(1+d/2)},\\
\label{eq:BQ_d5}
\frac{d B(Q^2)}{d m^2} &=& -2 \Lambda^{d-6} \hat B_m\left(\frac{Q^2}{\Lambda^2}, \frac{m^2}{\Lambda^2}\right),
\end{eqnarray}
where
\begin{eqnarray}
\hat B(q^2,\alpha^2) &=& \frac{d}{(4\pi)^{d/2}\Gamma(1+d/2)}\int_0^1  \rmd x\,\int_0^1 \frac{ y^{d-1}  \rmd y}{\left[y^2+\alpha^2 + q^2 x(1-x)\right]^2} , \\
\hat B_m(q^2,\alpha^2) &=& \frac{d}{(4\pi)^{d/2}\Gamma(1+d/2)}\int_0^1  \rmd x\,\int_0^1 \frac{ y^{d-1}  \rmd y}{\left[y^2+\alpha^2 + q^2 x(1-x)\right]^3} 
\end{eqnarray}
so that $B(Q^2) =O(\Lambda^{d-4})$.
As in the case at $d=4$, $u_r(s)=0$ in the limit that $\Lambda\rightarrow\infty$.

By the change of variable  $Q=\Lambda q$ in eq.~(\ref{eq:Z1}) and then taking the limit $\Lambda\rightarrow \infty$, we obtain
\begin{eqnarray}
Z_1 &= & \frac{2(4-d)}{d} \int_{q^2<1} \frac{\rmD q}{\hat B(q^2,0)}\frac{1}{(q^2)^2}.
\end{eqnarray} 
The fact that $\hat B(0,0)\not=0$ establishes that  $Z_1$ is finite at $d > 4$.

The most divergent part of $\mu_1^2$ is given by
\begin{eqnarray}
\mu_1^2 &\simeq& -\frac{u}{3} Z(m) \delta_1,
\end{eqnarray}
where
\begin{eqnarray}
\delta_1 &=& (d-2) \int_0^1 \frac{q^{d-1} \rmd q}{\hat B(q^2,0)} \left( \frac{\hat B(0,0)}{q^2} -\hat B_m(q^2,0)   \right)
\end{eqnarray}
with the change of variables as $q^2 = Q^2/\Lambda^2$. It is easy to show that $\delta_1$ is finite.

\section{Solving the SED for the flow equation}  
\label{app:SDE_flow}
In this appendix we explicitly solve the SDE in $d+1$ dimensions, in order to  obtain the 2-pt and 4-pt functions for the flow fields at the NLO.

\subsection{Solution for $\Gamma_0$}
We first solve the equation at the LO for $\Gamma_0$. If we introduce one unknown function $F(t,p)$ as
\begin{eqnarray}
\Gamma_0(12) &=& \int \rmD p \frac{F(t_1,p) F(t_2, p)}{p^2+m^2} \rme^{-(p^2+\mu_f^2)(t_1+t_2)}\rme^{ip(x_1-x_2)}
\end{eqnarray}
with the initial condition $F(0,p) =1$, we have
\begin{eqnarray}
D_1^f \, \Gamma_0(12) &=&  \int \rmD p \frac{\dot F(t_1,p) F(t_2, p)}{p^2+m^2} \rme^{-(p^2+\mu_f^2)(t_1+t_2)}\rme^{ip(x_1-x_2)} \\
-\frac{u_f}{6}\Gamma_0(12)\Gamma_0(11) &=& -\frac{u_f}{6}  \int \rmD p \frac{F(t_1,p) F(t_2, p)}{p^2+m^2} \rme^{-(p^2+\mu_f^2)(t_1+t_2)}\rme^{ip(x_1-x_2)}\Gamma_0(t_1), \\
\Gamma_0(t_1) &=& \Gamma_0(11)= \int \rmD p \frac{F^2(t_1,p)}{p^2+m^2} \rme^{-2(p^2+\mu_f^2)t_1} ,
\end{eqnarray}
where $\dot F$ means a  $t$-derivative of $F$.
Then, the SDE (\ref{eq:SDE_LO}) becomes
\begin{eqnarray}
\frac{\dot F(t,p)}{F(t,p)} &=& -\frac{u_f}{6} \Gamma_0(t) ,
\end{eqnarray}
which tells us that $F(t,p)$ is independent  of $p$,  so we put  $F(t,p)=F(t)$.  The above equation is thus reduced to
\begin{eqnarray}
\dot F(t) &=& - \frac{u_f}{6}F^3(t) \rme^{-2\mu_f^2 t} \zeta_0(t), 
\end{eqnarray}
  where $\zeta_0(t)$ is defined in eq.~(\ref{eq:zeta0}),
whose solution is given by 
\begin{eqnarray}
F^{-2}(t) = 1+ \frac{u_f}{3}\int_0^t \rmd s \zeta_0(s) \rme^{-2\mu_f^2 s} \equiv \rme^{-2\mu_f^2 t} \frac{\zeta(t)}{Z(m_f)}, \qquad \zeta(t) =\zeta_0(t) +\Delta(t)
\label{eq:zeta}
\end{eqnarray}
where  $m_f$ is defined in eq.~(\ref{eq:Zm})  and
\begin{eqnarray}
\Delta(t) &=& \rme^{2t \mu_f^2} \left( Z(m_f) - Z(m) \right) +\int \rmD p \left(\frac{p^2+m_f^2}{p^2+m^2}\right)\frac{\rme^{2t\mu_f^2} -\rme^{-2t p^2}}{p^2+\mu_f^2} .
\end{eqnarray}

In the case of the interacting flow with $u_f > 0$, $\mu_f^2$ negatively diverges as $Z(m_f)\rightarrow+\infty $ in the continuum limit at $d > 1$ or as $u_f\rightarrow +\infty$ in the NLSM limit.
In these limits, $\Delta(t)$ vanishes as 
\begin{eqnarray}
\lim_{\mu_f^2\rightarrow -\infty} \Delta(t) \simeq -\frac{m_f^2 \zeta_0(t) -\dot \zeta_0(t)/2}{\mu_f^2}+O\left(1/\mu_f^4\right) 
\end{eqnarray}
for $t > 0$. In the case of free flow ($u_f=0$), we simply have $F(t)=1$.

We then obtain
\begin{eqnarray}
\Gamma_0(12) &=& \left\{
\begin{array}{ll}
\displaystyle \dfrac{Z(m_f)}{\sqrt{\zeta(t_1)\zeta(t_2)} } \int \rmD p \dfrac{\rme^{-p^2(t_1+t_2)}\rme^{ip(x_1-x_2)}} {p^2+m^2}, & u_f \not= 0 \\
\\
\displaystyle \int \rmD p \dfrac{\rme^{-(p^2+\mu_f^2)(t_1+t_2)}\rme^{ip(x_1-x_2)}} {p^2+m^2}, & u_f=0 \\
\end{array}
\right. .
\end{eqnarray}

\subsection{Solution for $K_0$}
We consider $K_0$, which appears at the NLO.
The equation for $K_0$ in eq.~(\ref{eq:SDE_K0}) is closed, once $\Gamma_0$ is obtained.
Using eq.~(\ref{eq:K0g}),
we have
\begin{eqnarray}
D_1^f K_0(12;34) &=& \int  \rmd P_4 \left[ \frac{\dot F(t_1)}{F(t_1)} +\partial_{t_1}\right] g(12;34\vert 12;34)\nonumber \\
&=& \int  \rmd P_4 \left[ -\frac{u_f}{6} F^2(t_1)\rme^{-2\mu_f^2 t_1}\zeta_0(t_1) +\partial_{t_1}\right] g(12;34\vert 12;34), \\
\Gamma_0(12) \Gamma_0(13) \Gamma_0(14) &=& \int  \rmd P_4 \hat\delta\, (p_1^2+m^2) F^2(t_1) \rme^{-2\mu_f^2 t_1} \rme^{(p_1^2-p_2^2-p_3^2-p_4^2)t_1}, \\
\Gamma_0(11) K_0(12;34) &=& F^2(t_1) \rme^{-2\mu_f^2 t_1} \zeta_0(t_1)\int  \rmd P_4\, g(12;34\vert 12;34),\\
\Gamma_0(12) K_0(11;34) &=& F^2(t_1) \rme^{-2\mu_f^2 t_1} \int  \rmd P_4\hat\delta\, (p_1^2+m^2) \rme^{t_1(p_1^2-p_2^2)} \nonumber \\
&\times& \int \rmD q_1\rmD q_2\, \frac{\rme^{-t_1(q_1^2+q_2^2)}}{(q_1^2+m^2)(q_2^2+m^2)}\, g(11;34\vert q_1q_2;34), 
\end{eqnarray}
so that the SDE leads to
\begin{eqnarray}
\partial_{t_1} g(12;34\vert 12;34) &=& -\frac{u_f}{6}F(t_1)^2 \rme^{-2\mu_f^2 t_1} (p_1^2+m^2)\rme^{t_1(p_1^2-p_2^2)} \hat\delta \Bigl[ 2\rme^{-t_1(p_3^2+p_4^2)}\nonumber \\
&+& \int \rmD q_1\rmD q_2\, \frac{\rme^{-t_1(q_1^2+q_2^2)}}{(q_1^2+m^2)(q_2^2+m^2)}\, g(11;34\vert q_1q_2;34) \Bigr].
\label{eq:SDE_g}
\end{eqnarray}
 From eq.~(\ref{eq:SDE_g}),
one can easily see $\partial_{t_2}\partial_{t_1}g(12;34\vert 12;34) =0$, which implies
\begin{eqnarray}
g(12;34\vert 12;34) &=&X(23\vert 12;34) + X(13\vert 21;34)+X(24\vert 12;43)+X(14\vert 21;43)\nonumber \\
&+& Y(2\vert 12;34)+ Y(1\vert 21;34)+ Y(3\vert 43;12)+ Y(4\vert 34;12) \nonumber \\
&+& Z(\vert 12;34),
\end{eqnarray}
where we  require that $X$ and $Y$ satisfy  
\begin{eqnarray}
X(\tau,\tau^\prime \vert 12;34) &=& X(\tau^\prime,\tau \vert 43;21), \qquad
X(\tau,0 \vert 12;34) =0, \\
Y(\tau\vert 12;34) &=& Y(\tau\vert 12;43), \qquad Y(0 \vert 12;34) = 0.
\label{eq:Y}
\end{eqnarray}
Since $g(12;34\vert 12;34)$ agrees with the amputated connected 4-pt function in the $d$  dimensional theory at $ \tau_i = 0$ ($i=1,2,3,4$), we obtain
\begin{eqnarray}
Z(\vert p_1,p_2,p_3,p_4) &=& -\hat\delta \frac{2}{6/u+ B(0 \vert p_{34})},
\end{eqnarray}
 where $B(t \vert Q)$ is defined in eq.~(\ref{eq:Bt}).
Then one can easily check that  $g$ satisfies the required symmetries
\begin{eqnarray}
g(12;34\vert 12;34) &=& g(21;34\vert 21;34) =g(12;43\vert 12;43)=g(34;12\vert 34;12) .
\end{eqnarray}

\subsubsection{Solution  for $Y$}
Terms which  depend only on $t_1$ in eq.~(\ref{eq:SDE_g}) can be written as
\begin{eqnarray}
\partial_t Y(t\vert 21;34) &=& -\frac{u_f}{3} F^2(t) \rme^{-2t\mu_f^2}(p_1^2+m^2) \rme^{t(p_1^2-p_2^2)}
\hat\delta \nonumber \\
&\times& \left[ \rho(t\vert 34) +\int \rmD q_1\rmD q_2 \frac{\rme^{-t(q_1^2+q_2^2)} Y(t\vert q_1,q_2;34)}{(q_1^2+m^2)(q_2^2+m^2)}\right],
\label{eq:SDE_Y}
\end{eqnarray}
 where $\rho(t\vert34)$ is defined in eq.~(\ref{eq:rho_t}).
To solve this equation, we set
\begin{eqnarray}
Y(t\vert 21;34) &=& \hat\delta (p_1^2+m^2)\int_0^t \rmd s\, \rme^{s(p_1^2-p_2^2)}\psi(s\vert 34),
\end{eqnarray}
 satisfying eq.~(\ref{eq:Y}).
Eq.~(\ref{eq:SDE_Y}) is reduced to
\begin{eqnarray}
\psi(t\vert 34) &=& -\frac{u_f}{3} F^2(t)\rme^{-2t \mu_f^2} \left[\rho(t\vert 34) +\int_0^t \rmd s\, K(t,s\vert  p_{34}) \psi(s\vert 34) \right] ,
\end{eqnarray}
which shows $\psi$ does not depend on $p_1,p_2$,  where $K$ is defined in eq.~(\ref{eq:Kts}).
Since $u_f F^2(t) \rme^{-2t\mu_f^2} = u_f Z(m_f)/\zeta(t)$  goes to infinity in the continuum limit at
$t>0$ and  $d>1$ or in the NLSM   limit $u_f\rightarrow\infty$,   eq.~(\ref{eq:psi})
must hold 
in either of  the two limits.

\subsubsection{Solution  for $X$}
We next consider the solution for $X$. Terms depending on both $t_1$ and  $t_3$ in eq.~(\ref{eq:SDE_g}),
 and thereafter replacing $t_3$ by $t_2$ and interchanging $p_1\leftrightarrow p_2$,
 gives
\begin{eqnarray}
\partial_{t_1} X(t_1,t_2\vert 12;34) &=& -\frac{u_f}{6}F^2(t_1) \rme^{-2t_1\mu_f^2}(p_2^2+m^2) \rme^{t_1(p_2^2-p_1^2)} \hat\delta \int \rmD q_1\rmD q_2
\nonumber \\
&\times& \frac{\rme^{-t_1(q_1^2+q_2^2)}}{(q_1^2+m^2)(q_2^2+m^2)} \left\{ 2 X(t_1,t_2\vert q_1,q_2;34) +Y(t_2\vert 43;q_1,q_2) \right\},~~~
\end{eqnarray}
where
\begin{eqnarray}
Y(t\vert 43;q_1,q_2) &=&  (2\pi)^d \delta(p_{34}+q_{12}) (p_3^2+m^2) \int_0^t \rmd s\, \rme^{s(p_3^2-p_4^3)}\psi(s\vert q_1,q_2) .
\end{eqnarray}
We define
\begin{eqnarray}
\partial_{t_2}\partial_{t_1} X(t_1,t_2\vert 12;34) &=& \hat\delta (p_2^2+m^2) (p_3^2+m^2) \rme^{t_1(p_2^2-p_1^2)} \rme^{t_2(p_3^2-p_4^2)} \beta(t_1,t_2\vert 12;34),
\end{eqnarray}
where properties  of $X$ imply $\beta (t_1,t_2\vert 12;34) =\beta(t_2,t_1\vert 43;21)$ and $\beta(t,0\vert 12;34) =\beta(0,t\vert 12;34)=0$. 
Then the above equation becomes
\begin{eqnarray}
\beta(t_1,t_2\vert 12;34) &=& -\frac{u_f}{6}F^2(t_1) \rme^{-2t_1 \mu_f^2}\Bigl[g(t_1,t_2\vert p_{34}) + 2
\int_0^{t_1} \rmd s_1 \int \rmD q_1 \rmD q_2 (2\pi)^d\delta(q_{12}+p_{34})\nonumber \\
&\times& \frac{\rme^{-(t_1+s_1)q_1^2 -(t_1-s_1)q_2^2}}{q_1^2+m^2}\beta(s_1,t_2\vert q_1,q_2;34)\Bigr],
\end{eqnarray}
where
\begin{eqnarray}
g(t_1,t_2 \vert Q) = \int \rmD q_1\rmD q_2(2\pi)^d\delta(q_{12}+Q)\frac{\rme^{-t_1(q_1^2+q_2^2)}}{(q_1^2+m^2)(q_2^2+m^2)} \psi(t_2\vert q_1,q_2) .
\end{eqnarray}
Since the above expression tells us that $\beta$  depends only on $p_{34}$, we can write
\begin{eqnarray}
\beta(t_1,t_2\vert 12;34) &=& \omega(t_1,t_2\vert p_{34})=\omega(t_1,t_2\vert -p_{34}),
\end{eqnarray}
so that we have
\begin{eqnarray}
\omega(t_1,t_2\vert p_{34}) &=& -\frac{u_f}{6}F^2(t_1)\rme^{-2t_1 \mu_f^2}\Bigl[g(t_1,t_2\vert p_{34}) + 2\int_0^{t_1} \rmd s_1 K(t_1,s_1\vert p_{34}) \omega(s_1,t_2\vert p_{34})\Bigr], ~~~~~~~~
\end{eqnarray}
which is reduced to
\begin{eqnarray}
g(t_1,t_2\vert  Q) + 2\int_0^{t_1} \rmd s_1 K(t_1,s_1\vert  Q) \omega(s_1,t_2\vert  Q)
= 0 
\label{eq:g-omega}
\end{eqnarray}
in the continuum limit  or NLSM limit. 
Eq.~(\ref{eq:g-omega}) leads to eq.~(\ref{eq:omega}) in the main text, 
since
\begin{eqnarray}
\int_0^{t_2} ds_2\, K(t_2,s_2\vert Q) g(t_1,s_2\vert Q) &=& - \rho(t_1,t_2\vert Q).
\end{eqnarray}

\subsection{Solution for $\Gamma_1$}
\subsubsection{SDE at NLO}
The SDE for $\Gamma_1$ is a little  modified as
\begin{eqnarray}
D_1^f \Gamma_1(12) +\mu_{1,f}^2 \Gamma_0(12)&=& -\frac{u_f}{6}\Bigl[ K_0(12;11)+\Gamma_0(12)\Gamma_1(11) +\Gamma_1(12)\Gamma_0(11) + 2\Gamma_0(12)\Gamma_0(11)\Bigr], \nonumber \\
\end{eqnarray}
where we replace $\mu_f^2$ by $\mu_f^2+\dfrac{\mu_{1,f}^2}{N}$, so that
$D_1^f \rightarrow D_1^f +\dfrac{1}{N} \mu_{1,f}^2 $.
 Here $u_{1,f}^2$ is given by eq.~(\ref{eq:Z1_mu1}) with the replacement $u,m \rightarrow u_f, m_f$.

We parametrize $\Gamma_1$ as
\begin{eqnarray}
\Gamma_1(12) &=& F(t_1)F(t_2)\int \rmD p\frac{\rme^{-(p^2+\mu_f^2)(t_1+t_2)}\rme^{ip(x_1-x_2)}}{p^2+m^2} G_1(t_1,t_2\vert p) 
\end{eqnarray}
with the boundary condition 
\begin{eqnarray}
G_1(0,0\vert p) &\equiv& b(p) = -\frac{\Sigma_1(p)}{p^2+m^2} , 
\end{eqnarray}
where $\Sigma_1(p)$ is the self-energy at the NLO in the $d$  dimensional theory.

The NLO SDE becomes
\begin{eqnarray}
\partial_{t_1} G_1(t_1,t_2\vert  p_1) + \mu_{1,f}^2&=&   -\frac{u_f}{6}F^2(t_1)\rme^{-2t_1\mu_f^2}
{\cal H}\left[G_1(t_1,t_1|p)\right] + \lambda(t_1,t_2|p_1)\,,
\label{eq:G1}
\end{eqnarray}
where  ${\cal H}$ is defined in eq.~(\ref{eq:def_H})  and
\begin{eqnarray}
\lambda(t_1,t_2\vert p_1) &\equiv&
 -\frac{u_f}{6}F^2(t_1) \rme^{-2t_1\mu_f^2}\Delta(t_1,t_2\vert p_1), \\
\Delta(t_1,t_2\vert p_1) &=& 2 \zeta_0(t_1)
+\rme^{t_1p_1^2}\int \prod_{i=2}^4 
\frac{\rmD p_i \rme^{-t_1 p_i^2}}{p_i^2+m^2}\Bigl\{  Z(\vert 21;34) +2Y(1\vert 34;21) \nonumber \\
&+&2X(11\vert 12;34) +Y(1\vert12;34)+2X(21\vert 21;34) + Y(2\vert 21;34)\Bigr\} .
\end{eqnarray}

Using solutions $X$ and $Y$, we have  in the continuum limit
\begin{eqnarray}
\lambda(t_1,t_2\vert p_1) 
&=& \int \rmd p_2\, \frac{\rme^{t_1(p_1^2-p_2^2)}}{p_2^2+m^2}\Biggl[\psi(t_1\vert 12)+(p_2^2+m^2)\int_0^{t_1} \rmd  s\, \rme^{s(p_2^2-p_1^2)}\omega(t_1,s\vert p_{12}) \nonumber \\
&+&(p_1^2+m^2)\int_0^{t_2} \rmd s\, \rme^{s(p_1^2-p_2^2)}\omega(t_1,s\vert p_{12})\Biggr] .
\label{eq:lambda}
\end{eqnarray}
Since the right-hand side of eq.~(\ref{eq:lambda}) is finite,  $\Delta(t_1,t_2\vert p)\rightarrow 0$ in the continuum limit.

\subsubsection{Solution to the SDE}
\label{app:SDE_flow_G1}
Let us define
\begin{eqnarray}
G_1(t_1,t_2\vert p) &\equiv & b(p) + \kappa(t_1,t_2\vert p) + H(t_1) + H(t_2) 
\end{eqnarray}
with $\kappa(t_1,t_2\vert p) = \kappa(t_2,t_1\vert p)$  and $\kappa(0,0\vert p)=H(0)= 0$, where
\begin{eqnarray}
\partial_{t_1}\kappa(t_1,t_2\vert p) &=&\lambda(t_1,t_2\vert p), 
\label{eq:kappa}\\
\frac{d H(t)}{d t} &=& -\frac{u_f}{6}F^2(t)\rme^{-2t\mu_f^2}\left[
  {\cal H}\left[G_1(t,t\vert p)\right]
-2\zeta(t)\delta_1\right].
\label{eq:second}
\end{eqnarray}
The second equation (\ref{eq:second})  can be rewritten as
\begin{eqnarray}
\frac{d H(t)}{d t} &=&  -\frac{u_f}{6}F^2(t)\rme^{-2t\mu_f^2}\left[2\zeta_0(t) H(t) +b_0(t) +\kappa_0(t) -2\zeta(t) \delta_1\right],
\end{eqnarray}
so that we have  in the continuum limit
\begin{eqnarray}
H(t) &=& -\frac{b_0(t)+\kappa_0(t)}{2\zeta_0(t)} +\delta_1,
\end{eqnarray}
where we define $b_0(t) = {\cal H}[b(p)]$ and $ \kappa_0(t) ={\cal H}[\kappa(t,t\vert p)]$.

The first equation (\ref{eq:kappa}) can be solved as
\begin{eqnarray}
\kappa(t_1,t_2\vert p) &=& k_2(t_1,t_2\vert p) + k_1(t_1\vert p)+ k_1(t_2\vert p),
\end{eqnarray}
where
\begin{eqnarray}
k_1(t\vert p) &=& \int_0^t \rmd s\, \lambda_1(s\vert p), 
\label{eq:k1}\\
 \lambda_1(t\vert p) &=& 
 \int \rmD q \frac{\rme^{(p^2-q^2)t}}{q^2+m^2}\psi(t\vert p,q) +
 \int_0^t \rmd s \int \rmD q \rme^{(p^2-q^2)(t-s)} \omega(t,s\vert Q), 
\label{eq:lambda1} \\
 k_2(t_1,t_2\vert p) &=& \int_0^{t_1} \rmd s_1\int_0^{t_2} \rmd s_2 \int \rmD q \frac{p^2+m^2}{q^2+m^2} \rme^{(p^2-q^2)(s_1+s_2)}\omega(s_1,s_2\vert Q)
 \label{eq:k2}
\end{eqnarray}
with $Q=p+q$. 

\section{Calculations in the massless limit at $d=3$}
\label{app:massless}
 It can be shown that the flow bubble integral can be represented as
\begin{eqnarray}
B(t\vert  \{ p\}_{\rm sym.}) &=& -2\int_0^{ t} \rmd s\,  K (s,0 \vert   \{ p\}_{\rm sym.}) + B(0\vert   \{ p\}_{\rm sym.}), \quad B(0\vert   \{ p\}_{\rm sym.}) =\frac{1}{8\sqrt{D}},~~~~~~~
\end{eqnarray}
which can be rescaled as
\begin{eqnarray}
B(t\vert  \{ p\}_{\rm sym.}) &=& \frac{1}{\sqrt{D}}b_0(Dt),
\end{eqnarray}
where
\begin{eqnarray}
b_0(w) &=&\frac{1}{8} -\frac{\sqrt{w}}{2(2\pi)^{3/2}}\int_0^1\frac{ \rmd x}{\sqrt{x}}\rme^{-wx}\int_0^1\frac{ \rmd z}{\sqrt{z}}\rme^{wzx/2}.
\label{eq:b0}
\end{eqnarray}

Rescaling
\begin{eqnarray}
\rho(t\vert   \{ p\}_{\rm sym.}) &=& R_0(Dt, D),
\quad
\psi(t \vert   \{ p\}_{\rm sym.}) = \sqrt{D} \phi_0(Dt,D),
\end{eqnarray}
the integral equation for $\psi$ in the massless limit is written as
\begin{eqnarray}
R_0(w, D) +\int_0^w \rmd v\, k_0(w,v) \phi_0(v,D) &=& 0 ,
\end{eqnarray}
where
\begin{eqnarray}
R_0(w,D) &=& \rme^{-3w/2} -8b_0(w)\dfrac{\bar u(D)}{1+ \bar u(D)},
\qquad \bar u(D)= \dfrac{u}{48\sqrt{D}}.
\end{eqnarray}
Since the problem is linear, we can write
\begin{eqnarray}
\phi_0(w,D) &=& \phi_0^{(1)}(w) - 8 \phi_0^{(2)}(w) \frac{\bar u(D)}{1+\bar u(D)} ,
\end{eqnarray}
where $\phi_0^{(i)}$,  $i=1,2$ solve the momentum-independent equations (\ref{eq:phi0_1}) and (\ref{eq:phi0_2}).
We thus finally obtain eq.~(\ref{eq:hatY}).

As the source term can be rescaled as
\begin{eqnarray}
\rho(t,s\vert  \{ p\}_{\rm sym.}) &=&\frac{1}{\sqrt{D}}\left[ b_0(D(t+s)) -8 b_0(Dt) b_0(Ds)\frac{\bar u(D)}{1+\bar u(D)} \right], 
\end{eqnarray}
the equation for $\omega$ in the massless limit is written for $\omega(t,s\vert  \{ p\}_{\rm sym.})= \sqrt{D}W_0(Dt,Ds,D)$ as
\begin{eqnarray}
&&b_0(D(t+s)) -8 b_0(Dt) b_0(Ds)\frac{\bar u(D)}{1+\bar u(D)} \nonumber \\&=&
2\int_0^{Dt} \rmd u\, k_0(Dt,u) \int_0^{Ds} \rmd v\, k_0(Ds,v) W_0( u,v,D),
\end{eqnarray}
which can be solved as
\begin{eqnarray}
W_0(w,v , D) &=& \Omega_0(w,v) - 4 \phi_0^{(2)}(w) \phi_0^{(2)}(v) \frac{\bar u(D)} {1+\bar u(D)},
\end{eqnarray}
where $\Omega_0$ solves the momentum ($D$) independent equation (\ref{eq:Omega0}).
We thus obtain eq.~(\ref{eq:hatX}).

\section{Induced metric in the massless limit at $d=3$}
\label{app:metric}
\subsection{Induced metric}
The space component of the induced metric is given by
\begin{eqnarray}
g_{ij}(z) &=& \delta_{ij}\frac{R_0^2}{d\zeta_0(t)} \left(1-\frac{\zeta_1(t)}{N}\right) {\cal H}\left[ p^2\left(1 +\frac{G_1(t,t\vert p)}{N}\right)\right] .
\end{eqnarray}
We then evaluate
\begin{eqnarray}
\zeta_1(t) &=& \frac{1}{\zeta_0(t)} {\cal H}[ G_1(t,t\vert p)] = 2 \delta_1, \quad
{\cal H}[1] = \zeta_0(t), \qquad  {\cal H}[p^2] = -\frac{\partial_t\zeta_0(t)}{2}, \\ 
{\cal H}[ p^2 G_1(t,t\vert p)] &=& {\cal H}[\lambda(t,t\vert p)] +\zeta_0(t) \partial_t H(t) -\partial_t\zeta_0(t)\delta_1
=\zeta_0(t) \partial_t H(t) -\partial_t\zeta_0(t)\delta_1, ~~~~~
\end{eqnarray}
where in the last equation we use $ {\cal H}[\lambda(t,t\vert p)] = 0$. Altogether we obtain
\begin{eqnarray}
g_{ij}(z) &=& \delta_{ij} R_0^2\left[ g^{(0)}(t) + \frac{1}{N} g^{(1)}(t) \right], \quad
g^{(0)}(t) =-\frac{\partial_t\zeta_0(t)}{2d\zeta_0(t)}, \ g^{(1)}(t) =\frac{\partial_t H(t)}{d}.~~
\end{eqnarray}

The time component is evaluated as
\begin{eqnarray}
g_{00}(t)&=&t\partial_{t_1}\partial_{t_2}\left[
\frac{ R_0^2}{\sqrt{\zeta_0(t_1)\zeta_0(t_2)}}\int \rmD p\frac{{\mathrm e}^{-p^2(t_1+t_2)}}{p^2+m^2}
\left(1+\frac{\widetilde G_1(t_1,t_2|p)}{N}\right)\right]_{t_1=t_2=t}\\
&=& R_0^2\left\{ g_{00}^{(0)}(t) +\frac{1}{N}g_{00}^{(1)}(t)\right\},
\end{eqnarray} 
where
\begin{equation}
\widetilde G_1(t_1,t_2|p)=-2\delta_1+G_1(t_1,t_2|p).
\end{equation}
The leading term is
\begin{equation}
g_{00}^{(0)}(t)=\frac{t}{4}\partial_t^2\left[\log\zeta_0(t)\right]
\end{equation} 
and for the NLO term we have
\begin{equation}
\frac{1}{t}\,g^{(1)}_{00}(t)
=\partial_{t_1}\partial_{t_2}\frac{I(t_1,t_2)}{\sqrt{\zeta_0(t_1)\zeta_0(t_2)}}
\Bigg\vert_{t_1=t_2=t},
\end{equation} 
where
\begin{equation}
I(t_1,t_2)=
\int \rmD p\,\frac{{\mathrm e}^{-p^2(t_1+t_2)}}{p^2+m^2}\widetilde G_1(t_1,t_2|p).
\end{equation} 
With this notation
\begin{equation}
\frac{1}{t}\,g^{(1)}_{00}(t)=
\frac{1}{4}\,\frac{(\partial_t\zeta_0(t))^2}{\zeta_0^3(t)}\,I(t,t)
-\frac{1}{2}\,\frac{\partial_t\zeta_0(t)}{\zeta_0^2(t)}\,\partial_t I(t,t)
+\frac{1}{\zeta_0(t)}\partial_{t_1}\partial_{t_2} I(t_1,t_2)
\big\vert_{t_1=t_2=t}.
\end{equation} 
Since
\begin{equation}
I(t,t)={\cal H}[\widetilde G_1(t,t|p)]=0,
\end{equation} 
the first two terms vanish. Further,
\begin{equation}
\partial_{t_1}\partial_{t_2} I(t_1,t_2)\big\vert_{t_1=t_2=t}=
{\cal H}\big[(p^2)^2 \widetilde G_1(t,t|p)-2p^2\lambda(t,t|p)
+\partial_{t_2}\lambda(t,t_2|p)\big\vert_{t_2=t}\big]
+\partial_t H(t)\partial_t\zeta_0(t).
\end{equation} 
Using the identities
\begin{equation}
{\cal H}[\lambda(t,t|p)]=0;\qquad
{\cal H}[p^2\widetilde G_1(t,t|p)]=\zeta_0(t)\,\partial_t H(t)
\end{equation} 
and their derivatives this can be further simplified:
\begin{equation}
\partial_{t_1}\partial_{t_2} I(t_1,t_2)\big\vert_{t_1=t_2=t}=
-\frac{1}{2}\,\zeta_0(t)\partial_t^2 H(t)+
{\cal H}\big[\partial_{t_2}\lambda(t,t_2|p)\big\vert_{t_2=t}
-\,\partial_t\lambda(t,t|p)/2].
\end{equation} 
Here the second term vanishes  and we finally obtain
\begin{equation}
g^{(1)}_{00}(t)=-\frac{t}{2}\,\partial_t^2 H(t).
\end{equation}

\subsection{Calculation of $H(t)$ in the massless limit}
We recall the definition of $H(t)$ as
\begin{eqnarray}
H(t) &=& -\frac{b_0(t) +\kappa_0(t)}{2\zeta_0(t)} +\delta_1
\end{eqnarray}
where
\begin{eqnarray}
b_0(t)&=&{\cal H}[b(p)], \qquad \kappa_0(t) ={\cal H}[\kappa(t,t\vert p)],
\end{eqnarray}
with
\begin{eqnarray}
b(p) &=& -\frac{\Sigma_1(p)}{p^2+m^2}, \quad
\kappa(t,t\vert p) = k_2(t,t \vert p) + 2 k_1(t\vert p) .
\end{eqnarray}
 Here  $k_1$ and $k_2$ are given in eqs.~(\ref{eq:k1}), (\ref{eq:lambda1}) and (\ref{eq:k2}).

Hereafter we consider the massless limit at $d=3$, where we have
$\zeta_0(t)^{-1} = 2(2\pi)^{3/2} \sqrt{t}$.

\subsubsection{Calculation of $b_0(t)$}
We first calculate $b_0(t)$.  In the massless limit, we have
\begin{eqnarray}
H_b(t) &\equiv& -\frac{b_0(t)}{2\zeta_0(t)} =\frac{1}{2\zeta_0(t)}\int \rmD p \frac{\rme^{-2p^2 t}} {(p^2)^2} g(p^2) 
\end{eqnarray}
 since $\tilde C = Z_1 m^2 =0$ and
\begin{eqnarray}
g(p^2) &=&\frac{u}{3} \int \frac{\rmD Q}{1+\bar u(Q^2)}\left\{\frac{1}{(Q+p)^2} -\frac{1}{Q^2} \right\}.
\end{eqnarray}
After rescaling, we obtain 
\begin{eqnarray}
H_b(t) &=& \int \rmD Q\, h_b(Q^2) \frac{\bar u(Q^2) \sqrt{t}}{1+\bar u(Q^2) \sqrt{t}},
\end{eqnarray}
where
\begin{eqnarray}
h_b(Q^2) &=& 32\sqrt{2} \sqrt{\pi^3}\sqrt{Q^2} \int \rmD p\, \frac{\rme^{-2p^2}}{(p^2)^2} \left\{ \frac{1}{(Q+p)^2}-\frac{1}{Q^2}\right\} .
\end{eqnarray}

\subsubsection{Calculation of $\kappa_0(t)$}
For this we need $\psi$ and $\omega$ in the massless limit, which can be obtained as
\begin{eqnarray}
\psi_0(t\vert p,q) &=& \sqrt{Q^2}\left[ \varphi_0(Q^2t, z) - 8\phi_0^{(2)}(Q^2 t) \frac{\bar u(Q^2)}{1+\bar u(Q^2)} \right], \\
\omega_0(t,s \vert Q) &=& \sqrt{Q^2}\left[ \Omega_0(Q^2t,Q^2s) - 4\phi_0^{(2)}( Q^2t) \phi_0^{(2)}( Q^2s)
\frac{ \bar u(Q^2)}{1+\bar u(Q^2)} \right] 
\end{eqnarray}
with $z=(p^2+q^2)/Q^2$, where $\phi_0^{(2)}$ and $\Omega_0$ are already obtained in section~\ref{sec:coupling}, while
$\varphi_0$ satisfies
\begin{eqnarray}
\rme^{-z w} +\int_0^{w}  \rmd x\, k_0(w,x) \varphi_0(x,z) &=& 0,
\end{eqnarray}
 instead of eq.~(\ref{eq:phi0_1}) and thus $\varphi_0(x,3/2) =\phi_0^{(1)}(x) $.

Using these, we first calculate
\begin{eqnarray}
H_{\kappa}^{(1)}(t) &\equiv & -\frac{1}{\zeta_0(t)} \int \rmD p \frac{\rme^{-2p^2t}}{p^2} \int_0^t \rmd s\, \int \rmD q
\frac{\rme^{(p^2-q^2)s}}{q^2} \psi_0(s\vert p,q) \nonumber \\
&=& H_\kappa^{(1)}(0) +\int \rmD Q\, \int_0^1 \rmd x\,  \phi_0^{(2)}(Q^2 x) h_{11}(x,Q^2) \frac{ \bar u(Q^2)\sqrt{t}}{1+\bar u(Q^2) \sqrt{t}},
\end{eqnarray}
where  $H_\kappa^{(1)}(0)$ is some constant and
\begin{eqnarray}
h_{11}(x,Q^2) &=& 32\sqrt{2}\sqrt{\pi^3}\sqrt{Q^2}\int\rmD p\, \rmD q\, (2\pi)^3\delta(q+p-Q) \frac{\rme^{-(2-x)p^2-xq^2}}{p^2 q^2} .
\end{eqnarray}

Similarly we have
\begin{eqnarray}
H_{\kappa}^{(2)}(t) &\equiv & -\frac{1}{\zeta_0(t)} \int \rmD p \frac{\rme^{-2p^2t}}{p^2} \int_0^t \rmd s\, \int \rmD q\,
\rme^{(p^2-q^2)s} \int_0^s \rmd r\,  \rme^{(q^2-p^2)r}\omega_0(s,r\vert Q) \nonumber \\
&=& H_{\kappa}^{(2)}(0) + 2 \int \rmD Q\, \int_0^1  \rmd x\, \phi_0^{(2)}(Q^2x)  \int_0^x  \rmd y\, \phi_0^{(2)}(Q^2y) \,
h_{10}(x-y,Q^2) \nonumber \\
&\times & \frac{\bar u(Q^2)\sqrt{t}}{1+ \bar u(Q^2)\sqrt{t}},
\end{eqnarray}
where
\begin{eqnarray}
h_{10}(z,Q^2) &=& 8\sqrt{2}\sqrt{\pi^3}\sqrt{Q^2}\int \rmD p\, \rmD q\, (2\pi)^{3}\delta(q+p-Q) \frac{\rme^{-(2-z)p^2-z q^2}}{p^2} .~~~~
\end{eqnarray}

The last contribution becomes
\begin{eqnarray}
H_{\kappa}^{(3)}(t) &\equiv & -\frac{1}{2\zeta_0(t)} \int \rmD p\, \rme^{-2p^2t} \int_0^t \rmd s\, \int \rmD q\,
\frac{\rme^{(p^2-q^2)s}}{q^2} \int_0^t \rmd r\,  \rme^{(p^2-q^2)r}\omega_0(s,r\vert Q)\nonumber \\
&=&  H_{\kappa}^{(3)}(0) + \int \rmD Q\, \int_0^1  \rmd x\, \phi_0^{(2)}(Q^2x)  \int_0^1  \rmd y\, \phi_0^{(2)}(Q^2y) \, h_{10}(2-x-y,Q^2)
\nonumber \\
&\times & \frac{\bar u(Q^2)\sqrt{t}}{1+ \bar u(Q^2)\sqrt{t}}.
\end{eqnarray}

\subsection{Total contributions}
We thus obtain the $H(t)$ as\footnote{Here $H(0)$ is potentially divergent but it does not contribute to the metric.}
\begin{eqnarray}
H(t) &=& H(0) +\int \rmD Q\  h_{\rm total}(Q^2) \frac{\bar u(Q^2)\sqrt{t}}{1+ \bar u(Q^2)\sqrt{t}},
\end{eqnarray}
where
\begin{eqnarray}
H(0) &=& H_\kappa^{(1)}(0) +H_\kappa^{(2)}(0) + H_\kappa^{(3)}(0)+\delta_1 \\
h_{\rm total}(Q^2) &=& h_b(Q^2) + \int_0^1 \rmd x\, \phi_0^{(2)}(Q^2x) \left\{h_{11}(x,Q^2)
+2\int_0^x \rmd y\, \phi_0^{(2)}(Q^2y) h_{10}(x-y,Q^2) \right. \nonumber \\
&+&\left.
\int_0^1  \rmd y\, \phi_0^{(2)}(Q^2y)h_{10}(2-x-y,Q^2) \right\} ,
\end{eqnarray}
which leads to eqs.~(\ref{eq:A1}) and (\ref{eq:pA1}) by $A_1(t) \equiv \partial_t H(t)$ and $ \partial_t A_1(t) \equiv \partial^2_t  H(t)$.

\subsection{IR behaviors}
\subsubsection{Some definitions}
We write the NLO induced metric as
\begin{eqnarray}
g_{ij}(\tau) &=&\delta_{ij} \left\{ \frac{R_0^2}{12 t} \left[ 1+\frac{R(t)}{N} \right]\right\}, \qquad
g_{00}(\tau) = -t \partial_t \left\{ \frac{R_0^2}{8 t} \left[ 1+\frac{R(t)}{N} \right]\right\},
\end{eqnarray}
where the relative correction is a sum of four contributions,
\begin{equation}
R(t)=R_b(t)+\sum_{i=1}^3R^{(i)}_\kappa(t),\quad  R_b(t) \equiv 4t\partial_t H_b(t), \ R_\kappa^{(i)}(t) \equiv 4t\partial_t H_\kappa^{(i)}(t).
\end{equation}
We also introduce $G(v)$ by
\begin{equation}
\phi^{(2)}_{ 0}(v)=-\frac{(2\pi)^{3/2}}{\sqrt{v}}G(v),\qquad G(0)=1/8, \quad G(v)\sim \exp(-v/2),\  v\rightarrow\infty
\end{equation}
and use the time variable $T = u\sqrt{t}/48$.

 In the following we will use the fact that a
double 3-dimensional integral of any function depending only on the absolute
values $p$, $q$ and $|Q|$, where $Q=p+q$, can be written
\begin{equation}
  \int \rmD p\int \rmD q \, f(p,q,Q^2)=\frac{1}{(2\pi)^4}\int_0^\infty p \rmd p\, \int_0^\infty
q\rmd q\, \int_{(q-p)^2}^{(q+p)^2}\rmd Q^2\, f(p,q,Q^2) .
\end{equation}

\subsubsection{The $R_b$ contribution}
Here we can do the angular part of the $Q^2$ integral analytically and find
\begin{equation}
R_b(t)=\frac{32T}{\sqrt{2\pi}^5}\int_0^\infty\frac{q\rmd q}{(q+T)^2}\, \rho_b(q),  
\end{equation}
where
\begin{equation}
\rho_b(q)=q^2\int_0^\infty\frac{\rmd p}{p^3}\, \rme^{-2p^2}\left\{
\ln\frac{(p+q)^2}{(p-q)^2}-\frac{4p}{q}\right\},
\end{equation}
which behaves as $\rho_b(q)= O(q) $ for small $q$, while
\begin{equation}
\rho_b(q)\sim \frac{\sqrt{2\pi}}{3q}, 
\end{equation}
for large $q$. 
Thus we can establish that $R_b(t)=O(T)$ for small $t$, while 
for large $t$
\begin{equation}
r_b\equiv  R_b(\infty)=\frac{8}{3\pi^2}=0.27019. 
\end{equation}

\subsubsection{The $R_\kappa^{(1)}$ contribution}
We have
\begin{equation}
R^{(1)}_\kappa(t)=-32(2\pi)^3\int \rmD p\int \rmD q\int_0^1\frac{\rmd x}{\sqrt{x}}
\frac{\rme^{-p^2(2-x)-q^2x}}{p^2q^2}\frac{|Q|T}{(T+|Q|)^2}G(Q^2x).
\end{equation}
Doing the $q^2$ integral first and introducing $x=y^2$ we can rewrite it as
\begin{equation}
-\frac{32}{\pi}\int_0^\infty\frac{\rmd p}{p}\, \rme^{-2p^2}\int_0^\infty
\frac{Q^2T}{(T+Q)^2}\rmd Q\int_0^1\rmd y\, G(Q^2y^2)
\int_{(Q-p)^2}^{(Q+p)^2}\frac{\rme^{(p^2-q^2)y^2}}{q^2}\rmd q^2.
\end{equation}
After some further rescaling we get
\begin{equation}
R^{(1)}_\kappa(t)=-\frac{64}{\pi}\int_0^\infty\rmd Q\, \frac{QT}{(T+Q)^2}\,
\rho^{(1)}_\kappa(Q),  
\end{equation}
where
\begin{equation}
\rho^{(1)}_\kappa(Q)=\int_0^\infty\frac{\rmd p}{p}\rme^{-2p^2}
\int_0^Q\rmd z\, G(z^2)Y(\frac{p}{Q},z),
\end{equation}
\begin{equation}
Y(\varepsilon,z)=\int_{|1-\varepsilon|}^{1+\varepsilon}\frac{\rmd \xi}{\xi}\,
\rme^{(\varepsilon^2-\xi^2)z^2}=2\varepsilon\rme^{-z^2}+O(\varepsilon^2).
\end{equation}
From this we see that $\rho_\kappa^{(1)}(Q) = O(Q)$ for small $Q$, while 
\begin{equation}
\rho^{(1)}_\kappa(Q)\sim\frac{2}{Q}
\int_0^\infty\rmd p\, \rme^{-2p^2}\int_0^\infty
\rmd z\, G(z^2)
\rme^{-z^2}=\frac{1}{Q}\sqrt{\frac{\pi}{2}}\int_0^\infty\rmd z\,
G(z^2)\rme^{-z^2}      
\end{equation}
 for large $Q$,  so that  we numerically obtain 
\begin{equation}
r^{(1)}_\kappa \equiv R^{(1)}_\kappa(\infty)=-\frac{64}{\sqrt{2\pi}}\int_0^\infty\rmd z\, G(z^2)
\rme^{-z^2}=-1.14734.
\end{equation}

\subsubsection{The $R_\kappa^{(2)}$ contribution}
Similarly
\begin{equation}
R^{(2)}_\kappa(t)=16\sqrt{2\pi}^9\int \rmD p\int \rmD q\int_0^1
\frac{\rmd x}{\sqrt{x}}\int_0^x\frac{\rmd y}{\sqrt{y}}
\frac{T}{(T+|Q|)^2}G(Q^2x)G(Q^2y)
\frac{\rme^{-2p^2}}{p^2}
\rme^{(p^2-q^2)(x-y)}. 
\end{equation}
Doing the $q^2$ integrations first, we have
\begin{equation}
\begin{split}
R^{(2)}_\kappa(t)=64\sqrt{2\pi}&\int_0^\infty\rmd Q \frac{QT}{(T+Q)^2}
\int_0^\infty\frac{\rmd p}{p}\rme^{-2p^2}\\
&\times \int_0^1\rmd x\int_0^x\rmd y \, G(Q^2x^2) \, G(Q^2y^2)
\int_{(Q-p)^2}^{(Q+p)^2} \rme^{(p^2-q^2)(x^2-y^2)}\rmd q^2.
\end{split}
\end{equation}
The $q^2$ integral can be done analytically and we find
\begin{equation}
R^{(2)}_\kappa(t)=128\sqrt{2\pi}\int_0^\infty\rmd Q \frac{QT}{(T+Q)^2}
\rho^{(2)}_\kappa(Q),
\end{equation}
where
\begin{equation}
\rho^{(2)}_\kappa(Q)=
\int_0^\infty\frac{\rmd p}{p}\rme^{-2p^2}
\int_0^Q\rmd z\int_0^z\rmd w \, G(z^2) \, G(w^2)
\frac{\rme^{w^2-z^2}}{z^2-w^2}\sinh\frac{2p}{Q}(z^2-w^2).
\end{equation}
Thus  $\rho_\kappa^{(2)} = O(Q)$ for small $Q$, while
\begin{equation}
\rho^{(2)}_\kappa(Q)\sim\frac{1}{Q}
\int_{-\infty}^\infty\rmd p\, \rme^{-2p^2}
\int_0^\infty\rmd z\int_0^z\rmd w \, G(z^2) \, G(w^2)
\rme^{w^2-z^2}
\end{equation}
for large $Q$, and
\begin{equation}
r^{(2)}_\kappa \equiv R_\kappa^{(2)}(\infty) =
128\pi
\int_0^\infty\rmd z\int_0^z\rmd w \, G(z^2)\, G(w^2)
\rme^{w^2-z^2}=0.45846.
\end{equation}

\subsubsection{The $R_\kappa^{(3)}$ contribution}
For $R_\kappa^{(3)}$ we find
\begin{equation}
R_\kappa^{(3)}(t)=32\sqrt{2\pi}\int_0^\infty\rmd Q\frac{QT}{(T+Q)^2}
\rho^{(3)}_\kappa(Q)
\end{equation}
with
\begin{equation}
\rho^{(3)}_\kappa(Q)=\int_0^1\rmd x \int_0^1\rmd y
\int_0^\infty p\rmd p\, \rme^{-2p^2+p^2(x^2+y^2)}G(Q^2x^2)G(Q^2y^2)
\int_{(Q-p)^2}^{(Q+p)^2}\frac{\rme^{-q^2(x^2+y^2)}}{q^2} \rmd q^2.
\end{equation}
After rescaling
\begin{equation}
\rho^{(3)}_\kappa(Q)=\frac{1}{Q^2}\int_0^Q\rmd z \int_0^Q\rmd w\,
G(z^2)\, G(w^2)\int_0^\infty p\rmd p\, \rme^{-2p^2}
Z\left(\frac{p}{Q},z^2+w^2\right),
\end{equation}
where
\begin{equation}
Z(\varepsilon,A)=2\rme^{A\varepsilon^2}\int_{|1-\varepsilon|}^{1+\varepsilon}
\frac{\rme^{-A\xi^2}}{\xi}\rmd \xi
\approx 4\varepsilon\rme^{-A}, \quad \varepsilon\rightarrow 0.
\end{equation}
Thus $\rho_\kappa^{(3)}(Q) = O(Q)$ for small $Q$, while 
\begin{equation}
\begin{split}
\rho^{(3)}(Q)\sim\frac{4}{Q^3}\int_0^\infty\rmd z &\int_0^\infty\rmd w\,
G(z^2)\, G(w^2)\int_0^\infty p^2\rmd p\, \rme^{-2p^2-z^2-w^2}\\
&=\sqrt{\frac{\pi}{8}}\left(\int_0^\infty\rmd z\, G(z^2)\rme^{-z^2}\right)^2
\frac{1}{Q^3},
\end{split}
\end{equation}
 for large $Q$, which leads to 
\begin{equation}
r^{(3)}_\kappa \equiv R^{(3)}_\kappa(\infty)=0 . 
\end{equation}

Thus the total relative correction is negative:
\begin{equation}
r=r_b+r^{(1)}_\kappa+r^{(2)}_\kappa+r^{(3)}_\kappa=
-0.41869.
\end{equation}


\end{document}